\documentclass[prb,aps,tightenlines,twocolumn,floatfix,superscriptaddress,showpacs,preprintnumbers,citeautoscript,10pt]{revtex4-1}

\usepackage{amsmath}
\usepackage{graphicx,multirow,dcolumn}
\usepackage{float,array}
\usepackage{tikz}
\usepackage{bbm}
\usepackage{algorithm}
\usepackage{algpseudocode}
\usepackage{braket}
\usepackage{xfrac}
\usepackage[version=4]{mhchem}

\newcommand{\ie}[0]{\textit{i.e.}, }
\newcommand{\eg}[0]{\textit{e.g.}, }
\newcommand{\via}[0]{\textit{via} }
\newcommand{\cf}[0]{\textit{cf}. }
\usepackage{xspace}
\newcommand{\SCDMM}[0]{SCDM-M\xspace}
\newcommand{\SCDML}[0]{SCDM-L\xspace}
\newcommand{\SCDMG}[0]{SCDM-G\xspace}
\newcommand{\DefineAuthor}[2]{%
  \expandafter\newcommand\csname #1note\endcsname[1]{%
    \textbf{\textcolor{#2}{\textbf{#1:} ##1}}}%
  \expandafter\newcommand\csname #1\endcsname[1]{
    \textbf{\textcolor{#2}{##1}}}
  \expandafter\newcommand\csname #1cancel\endcsname[1]{%
    \textbf{\textcolor{#2}{\sout{##1}}}}%
  \expandafter\newcommand\csname #1change\endcsname[2]{%
    \textbf{\textcolor{#2}{\sout{##1} ##2}}}%
  \newenvironment{#1text}{\color{#2}}{\color{black}}
}
\definecolor{dartmouthgreen}{rgb}{0.05, 0.5, 0.06}
\DefineAuthor{RAD}{red}
\DefineAuthor{AD}{blue}
\DefineAuthor{EGF}{magenta}
\DefineAuthor{GEN}{dartmouthgreen}
\usepackage{soul}

\begin{document}

\title{Selected Columns of the Density Matrix in an Atomic Orbital Basis I: An Intrinsic and Non-Iterative Orbital Localization Scheme for the Occupied Space}
\author{Eric G. Fuemmeler}\thanks{Work completed while at Cornell University.}
\affiliation{Mayo Clinic, Rochester, MN 55905}
\author{Anil Damle}
\altaffiliation{E-mail: damle@cornell.edu}
\affiliation{Department of Computer Science, Cornell University, Ithaca, NY 14853}
\author{Robert A. DiStasio Jr.}
\altaffiliation{E-mail: distasio@cornell.edu}
\affiliation{Department of Chemistry and Chemical Biology, Cornell University, Ithaca, NY 14853}

\begin{abstract}
\noindent In this work, we extend the selected columns of the density matrix (SCDM) methodology [\textit{J. Chem. Theory Comput.}\ \textbf{2015}, \textit{11}, 1463--1469]---a non-iterative and real-space procedure for generating localized occupied orbitals for condensed-phase systems---to the construction of local molecular orbitals (LMOs) in systems described using non-orthogonal atomic orbital (AO) basis sets.
In particular, we introduce three different theoretical and algorithmic variants of SCDM  (referred to as \SCDMM, \SCDML, and \SCDMG) that can be used in conjunction with the AO basis sets used in standard quantum chemistry codebases.
The \SCDMM and \SCDML variants are based on a pivoted QR factorization of the Mulliken and L{\"o}wdin representations of the density matrix, and are tantamount to selecting a well-conditioned set of projected atomic orbitals (PAOs) and projected (symmetrically-) orthogonalized atomic orbitals (POAOs), respectively, as proto-LMOs that can be orthogonalized to exactly span the occupied space.
The \SCDMG variant is based on a real-space (grid) representation of the wavefunction, and therefore has the added flexibility of considering a large number of grid points (or $\delta$ functions) when selecting a set of well-conditioned proto-LMOs.
A detailed comparative analysis across molecular systems of varying size, dimensionality, and saturation level reveals that the LMOs generated by these three non-iterative/direct SCDM variants are robust, comparable in orbital locality to those produced with the iterative Boys or Pipek-Mezey (PM) localization schemes, and completely agnostic towards any single orbital locality metric.
Although all three SCDM variants are based on the density matrix, we find that the character of the generated LMOs can differ significantly between \SCDMM, \SCDML, and \SCDMG.
In this regard, only the grid-based \SCDMG procedure (like PM) generates LMOs that qualitatively preserve $\sigma\text{-}\pi$ symmetry (in systems like s-\textit{trans} alkenes), and are well-aligned with chemical (\ie Lewis structure) intuition.
While the direct and standalone use of SCDM-generated LMOs should suffice for most chemical applications, our findings also suggest that the use of these orbitals as an unbiased and cost-effective (initial) guess also has the potential to improve the convergence of iterative orbital localization schemes, in particular for large-scale and/or pathological molecular systems.
\end{abstract}

\maketitle

\section{Introduction \label{sec:intro}}

Localized molecular orbitals (LMOs) have proven to be an incredibly valuable tool in quantum chemistry since their introduction by Coulson in 1942 to study the nature of the C--H bond in methane,~\cite{Cou42} with further extensions by Lennard-Jones and Pople throughout the late 1940s~\cite{len49} and early 1950s.~\cite{len50,len51} 
Since then, LMOs have been used to bridge the gap between high-level quantum mechanical calculations and traditional concepts in chemistry,~\cite{hof71} such as bonding and lone pairs, orbital hybridization, and bond order.~\cite{tri69,cio92}
For example, LMOs have been used to elucidate the complex bonding nature of boranes,~\cite{Swi70,Eps73,Lip73,Kle74,Lip77} which often exhibit three-center two-electron bonds, as well as the interesting case of the ``inverted" bond seen in [1.1.1]propellane.~\cite{New72,Ebr03,Wu09p}
More recently, LMOs have also been used to quantum mechanically justify the ``curly arrow" notation used to denote the movement of electrons when discussing chemical reactions.~\cite{kni15,Vid17,gle19}

In addition to their analytical utility, LMOs can also be used to reduce the computational cost associated with wavefunction theory (WFT)~\cite{Sae87,sae88,Sae93,Mas98,rau98,Sch99,lee00b,wer03,dis05,Kat08,ham96,SCH00,Sch01,sch02,nee09,nee09b,hel11,rip13} and density functional theory (DFT)~\cite{Sol02,chen_x-ray_2010,hin11,swartz_ab_2013,Mus15,hes17,Wu09,gyg09,gyg13,dis14,daw15,hsi19a} methods, thereby providing an avenue towards linear-scaling algorithms~\cite{RGoe99} to treat large-scale systems.
This follows from the invariance of the mean-field Hartree-Fock (HF) and Kohn-Sham (KS) ground-state energies with respect to unitary (orthogonal) transformations of the occupied canonical molecular orbitals (CMOs),~\cite{Fock30} \ie the eigenfunctions of the Fock (or effective Hamiltonian) matrix.
With the freedom to work in a localized (instead of delocalized) representation of the occupied space, one can exploit the inherent sparsity in the quantum mechanical descriptions of molecules and materials.~\cite{Kohn96,Pro05}
For example, the use of occupied LMOs can greatly reduce the computational effort required to evaluate the exact exchange interaction, thereby enabling linear scaling calculations at the HF and hybrid DFT levels of theory.~\cite{Wu09,gyg09,gyg13,dis14,daw15,hsi19a}
Occupied (and virtual) LMOs are also at the very heart of local electron correlation (\eg post-HF) methods, which include M{\o}ller-Plesset perturbation theory,~\cite{Sae87,sae88,Sae93,rau98,Mas98,Sch99,lee00b,wer03,dis05,Kat08} coupled cluster theory,~\cite{ham96,SCH00,Sch01,sch02} as well as methods based upon pair natural orbitals.~\cite{nee09,nee09b,hel11,rip13}
In particular, the use of LMOs allows these methods to define and take advantage of local correlation domains, which significantly decrease the number of excitations that must be considered to capture subtle electron correlation effects.
In addition, LMOs can also be used in post-DFT methods such as the random phase approximation (RPA)~\cite{Mus15,hes17} and GW-based approaches,~\cite{chen_x-ray_2010,swartz_ab_2013} which explicitly account for higher-order many-body effects at a fraction of the computational cost.

In the field of computational molecular quantum mechanics, the most widely used occupied LMOs are arguably those introduced by Foster and Boys in the 1960s.~\cite{Fos60}
In this scheme, the unitary transformation from CMOs ($\{ \psi_i \}$) to LMOs ($\{ \phi_i \}$) \via $\ket{\phi_i} = \sum_j \ket{\psi_j} U_{ji}$ is accomplished by an iterative optimization (minimization) of the following cost function,~\cite{Boy66} 
\begin{align}
    \Omega_{\rm Boys} &= \sum_i^{N_{\rm occ}} \braket{\phi_i | \left( \hat {\mathbf{r}}-\braket{\phi_i|\hat {\mathbf{r}}|\phi_i} \right)^2 | \phi_i} \equiv \sum_i^{N_{\rm occ}}\sigma^2_i ,
    \label{eq:boys}
\end{align}
in which the sum over $i$ includes all $N_{\rm occ}$ occupied orbitals.
In doing so, the Foster-Boys (or simply Boys) procedure seeks to minimize the total spread functional, \ie the sum of the orbital variances ($\sigma_{i}^{2}$, or second central moments) associated with each LMO.
Orbital optimization with respect to this statistical measure of locality has proven to be an extremely powerful tool for generating LMOs, as evidenced by the wide number of applications found in both molecular systems and condensed-phase materials (\textit{vide infra}).~\cite{Wan37,mar97,Mar12} 
Other popular and historically relevant methods for generating LMOs in molecular systems include that of Pipek-Mezey (PM)~\cite{Pip89} and Edmiston-Ruedenberg (ER).~\cite{er63}
In the PM scheme, one seeks to iteratively maximize the following population-based cost function,~\cite{Pip89,hoy13,Leh14}
\begin{align}
    \Omega_{\rm PM} = \sum_A^{N_{\rm atom}} \sum_i^{N_{\rm occ}} |q_i^A|^2,
    \label{eq:pm}
\end{align}
in which $q_i^A$ is the electronic population on atom $A$ due to orbital $i$ (traditionally taken to be the population metrics given by Mulliken~\cite{mul55} or L{\"o}wdin~\cite{lowdin1950nonorthogonality}), and the sum over $A$ includes all $N_{\rm atom}$ atoms.
In the ER scheme, one seeks to iteratively maximize the total self-Coulombic repulsion of the LMOs \via the following metric,~\cite{er63}
\begin{align}
    \Omega_{\rm ER} = \sum_i^{N_{\rm occ}} (ii|ii) , 
    \label{eq:er}
\end{align}
in which $(ii|ii)$ are the two-electron repulsion integrals defined as:
\begin{align}
    (ii|ii) \equiv \int d\mathbf{r} \int d\mathbf{r'} \, \frac{\phi_i^{\ast}(\mathbf{r})\phi_i(\mathbf{r})\phi_i^{\ast}(\mathbf{r'})\phi_i(\mathbf{r'})}{|\mathbf{r}-\mathbf{r'}|} .
    \label{eq:er2}
\end{align}
Although the ER scheme was one of the earliest LMO methods developed, the steep computational cost associated with the assembly of the two-electron integrals in Eq.~\eqref{eq:er2} has limited its use in practice, despite algorithmic advances which asymptotically reduce the overall computational scaling.~\cite{sub04}
On the other hand, the PM method has enjoyed significantly more widespread use due to its favorable computational scaling and intrinsic preservation of $\sigma\text{-}\pi$ separation in molecules containing multiple (\ie double and triple) bonds and/or multiple lone pairs.~\cite{Pip89}
Despite producing otherwise similar LMOs, we note in passing that the Boys and ER methods tend to mix $\sigma$ and $\pi$ bonds during the localization procedure,~\cite{Pip89} thereby producing so-called $\tau$ or banana bonds;~\cite{pal86} these LMOs do not take molecular symmetry into account, but can be rationalized within the context of the equivalent-orbital model of valence bond theory.~\cite{pau31}
Although LMOs belonging to the $\sigma\text{-}\pi$ and equivalent-orbital representations are simply related by a unitary (or orthogonal) transformation and will yield the the same total electron density (and hence the same HF or KS energy), their use can lead to markedly different qualitative and quantitative interpretations when analyzing chemical systems.~\cite{pal86,win87,lai87,kar93,wib96,car10,cla14}

More recently, J{\o}rgensen and co-workers~\cite{Jan11,Hoy12} have explored several alternative metrics that extend the statistical measure of locality used in the Boys scheme (\cf $\Omega_{\rm Boys}$ in Eq.~\eqref{eq:boys}) to include higher (\eg fourth) central moments as well as powers of central moments; in doing so, these metrics effectively decrease the ``tails'' and increase the ``bulk'' locality of the LMOs, and have proven to be powerful approaches for generating highly localized orbitals.~\cite{Jan11,Hoy12} 
Although the popular (but ill-defined) Mulliken~\cite{mul55} and L{\"o}wdin~\cite{lowdin1950nonorthogonality} population metrics have been commonly employed in the definition of $\Omega_{\rm PM}$ in Eq.~\eqref{eq:pm}, other metrics (\eg Bader,~\cite{bad90} Hirshfeld,~\cite{hir77} Becke,~\cite{bec88} ``fuzzy atom''~\cite{alc06}) have also been the subject of recent interest.~\cite{Leh14}
In the same breath, Knizia~\cite{kni13} has also explored an extension of the population-based PM scheme \via the use of intrinsic atomic orbitals (IAOs), which are conceptually similar to the extracted polarized atomic orbitals (EPAOs) of Lee and Head-Gordon.~\cite{lee00}
Here, we also note that several direct (non-iterative) orbital localization schemes have also been proposed,~\cite{Aqu06,hes16} which avoid explicit optimization procedures and hold promise for generating LMOs for large-scale and/or pathological molecular systems.

Furthermore, the use of localized orbitals (LOs, not necessarily localized \textit{molecular} orbitals (LMOs)) has also been extended to the study of condensed-phase systems such as solids and liquids.
In order to construct a localized representation of the occupied KS orbitals (bands), which are typically expanded in a basis set of planewaves, Marzari and Vanderbilt have introduced the maximally localized Wannier function (MLWF) method,~\cite{Wan37,mar97,Mar12} which can be seen as the reciprocal-space analog of the Boys localization scheme discussed above.
In the same vein, J{\'o}nsson \textit{et al.} recently extended the Pipek-Mezey formalism described above to the study of extended systems.~\cite{jon17}
Over the past decade, Gygi and co-workers~\cite{gyg09,gyg13} have developed the so-called recursive subspace bisection (RSB) method, which is a novel orbital localization scheme that algebraically decomposes the wavefunction coefficients and transforms the KS eigenstates to a set of LOs contained within prescribed domains in real space.
As an alternative to the condensed-phase MLWF and RSB orbital localization schemes, Damle, Lin, and Ying~\cite{Dam15,Dam17,Dam17b,Dam18} have developed the selected columns of the density matrix (SCDM) approach, in which columns of the one-particle density matrix are used as templates (or proto-LOs) for a set of LOs that exactly span the occupied space.
In doing so, these proto-LOs inherit the well-known local structure (sparsity) of the density matrix,~\cite{Kohn96,Pro05} and therefore provide a natural starting point for generating LOs.
One benefit of the SCDM approach over methods like MLWFs, is that SCDM (like RSB) is a direct and non-iterative orbital localization scheme that does not require an initial guess for the LOs; as such, SCDM largely sidesteps issues related to convergence and dynamical gauge fields,~\cite{iftimie_on-the-fly_2004,thomas_field_2004} and therefore has the potential to furnish robust LOs for large-scale condensed-phase systems.

In this work, we extend the SCDM methodology to the non-iterative construction of LMOs in systems described using non-orthogonal and atom-centered (\ie atomic orbital, AO) basis sets---the standard framework for performing high-level quantum mechanical calculations on molecules and molecular systems.
We begin by reviewing the theoretical underpinnings of the SCDM method in the treatment of condensed-phase systems in Sec.~\ref{sec:review}, before introducing three new theoretical and algorithmic variants of SCDM  (\SCDMM, \SCDML, and \SCDMG) that can be used in conjunction with the AO basis sets used in standard quantum chemistry codebases in Sec.~\ref{sec:extension}.
We then evaluate the performance of these SCDM variants in Sec.~\ref{sec:rd} for molecular systems of varying size, dimensionality, and saturation level; in doing so, we provide a detailed comparative analysis that includes the popular Boys and PM iterative orbital localization schemes, and focuses on orbital compactness (locality) and chemical interpretability of the resulting LMOs. 
Our findings demonstrate that the LMOs generated by the three non-iterative/direct SCDM variants introduced herein are robust, comparable in orbital locality to those produced with the iterative Boys or PM localization schemes, and completely agnostic towards any single locality metric.
While the direct and standalone use of SCDM-generated LMOs should suffice for most chemical applications, we also briefly explore the use of these orbitals as an unbiased and cost-effective (initial) guess to improve the convergence of iterative orbital localization schemes in Sec.~\ref{sec:guess}.

\section{Theory \label{sec:theory}}

\subsection{Notation and Index Conventions \label{sec:notation}}

We will utilize the following conventions for the various symbols, dressings, and indices encountered in this work:
\begin{itemize}
\item $\chi$: atomic orbitals (AOs)
\item $\chi'$: orthogonalized atomic orbitals (OAOs)
\item $\psi$: canonical (molecular) orbitals, COs (CMOs)
\item $\phi$: localized (molecular) orbitals, LOs (LMOs)
\item $\mu,\nu,\lambda,\sigma$: indices for AOs and OAOs
\item $i,j,k$: indices for COs (CMOs) and LOs (LMOs)
\item $p$, $q$: general summation indices
\item $\mathbf{\Psi}$: COs (CMOs) represented on real-space grid
\item $\mathbf{\Phi}$: LOs (LMOs) represented on real-space grid
\item Symbols dressed with a tilde ($A\Rightarrow\widetilde{A}$) represent projected quantities (\ie proto-LOs/proto-LMOs)
\item $N_{\rm AO}$: number of AOs
\item $N_{\rm occ}$: number of \textit{active} occupied orbitals
\item $N_{\rm atom}$: number of atoms
\item $N_{\rm grid}$: number of real-space grid points
\end{itemize}

\subsection{Review of SCDM in Condensed-Phase Quantum Mechanics \label{sec:review}}

The ability to construct LOs using the SCDM procedure relies upon the underlying locality of the (one-particle) density matrix ($\mathbf{P}$).
When expressed in a local representation (\eg on a real-space grid), $\mathbf{P}$ will exhibit exponential decay~\cite{Kohn96,Pro05} away from the diagonal for insulating condensed-phase systems (\ie systems with a sizeable gap). 
As such, $\mathbf{P}$ itself can serve as a useful starting point for the construction of LOs, and it is this observation that forms the foundation of the SCDM approach.~\cite{Dam15,Dam17,Dam17b,Dam18}

We start by defining the set of $N_{\rm occ}$ occupied COs obtained by solving the KS (or HF) equations as $\{\psi_i(\mathbf{r})\}_{i=1}^{N_{\rm occ}}$, which satisfy the following orthonormality condition, $\braket{\psi_i|\psi_j}=\delta_{ij}$.
Without loss of generality, we will focus our discussion on the closed-shell (spin-unpolarized) case; as such, the contributions from the $\alpha$- and $\beta$-spin manifolds are equivalent, and we will only consider quantities corresponding to a single spin manifold (\eg $\alpha$) throughout this work.
Letting $\mathbf{\Psi}$ be an $N_{\rm grid} \times N_{\rm occ}$ matrix representing the COs in real space (with $N_{\rm grid}$ being the number of real-space grid points), the corresponding $N_{\rm grid} \times N_{\rm grid}$ density matrix is given by:
\begin{align}
    \mathbf{P}=\mathbf{\Psi}\mathbf{\Psi}^{\ast} .
    \label{eq:dm}
\end{align} 
Since every column of $\mathbf{P}$ is localized in real space, there are many ways to choose $N_{\rm occ}$ linearly independent columns of $\mathbf{P}$ that could serve as templates (or proto-LOs) when constructing a set of orthogonal and well-localized orbitals that exactly spans the occupied space.
However, not all of these choices will be equally effective; for instance, ensuring that the proto-LOs remain sufficiently local during the orthogonalization (or orthonormalization) procedure requires that the chosen columns of $\mathbf{P}$ should be suitably distinct from one another.

To deal with this issue, the SCDM procedure looks for a well-conditioned subset of columns of $\mathbf{P}$ to serve as proto-LOs.
We denote this subset of columns by $\mathcal{C}$, in which $\mathcal{C} \subset \{1,\ldots,N_{\rm grid}\}$ and $\lvert \mathcal{C} \rvert = N_{\rm occ}$.
To ensure that $\mathcal{C}$ is well-conditioned, a rank-revealing QR factorization~\cite{Bus65,che05,Gu96} is employed.
More specifically, a QR factorization with column pivoting~\cite{Note1} is applied to $\mathbf{P}$, \ie
\begin{align}
    \mathbf{P}\mathbf{\Pi}=\mathbf{Q}\mathbf{R} ,
    \label{eq:qr}
\end{align}
in which $\mathbf{\Pi}$ is a permutation matrix, $\mathbf{Q}$ is an orthogonal matrix, and $\mathbf{R}$ is an upper triangular matrix (all of which are $N_{\rm grid} \times N_{\rm grid}$).
As part of the QR factorization, $\mathbf{\Pi}$ (which encodes $\mathcal{C}$) is chosen such that the principle sub-matrices of $\mathbf{R}$ are as well-conditioned as possible and the diagonal elements of $\mathbf{R}$ are non-increasing, \ie $|R_{11}| \geq |R_{22}| \geq \cdots \geq |R_{N_{\rm grid}N_{\rm grid}}|$.
With this permutation matrix in hand, $\mathcal{C}$ is formally given by the row positions of the non-zero entries in the leading $N_{\rm occ}$ columns of $\mathbf{\Pi}$.
In practice, $\mathbf{\Pi}$ is represented as a permutation vector, and one simply takes the leading $N_{\rm occ}$ entries as $\mathcal{C}$.
We note in passing that QR factorization with pivoting is often used to determine the numerical rank of a matrix (or to compute lower-rank approximations to a given matrix).
For our purposes here (in which the rank of $\mathbf{P}$ is known to be exactly equal to $N_{\rm occ}$), the pivoted QR procedure is simply employed to identify a well-conditioned set of proto-LOs (\ie selected columns of $\mathbf{P}$) \via the computation of $\mathcal{C}$.

With $\mathcal{C}$ in hand, the $N_{\rm grid} \times N_{\rm occ}$ $\mathbf{\widetilde{\Phi}}$ matrix, which contains the set of proto-LOs ($\{{\widetilde{\phi}}_{i}(\mathbf{r})\}_{i=1}^{N_{\rm occ}}$) on the real-space grid, can then be extracted from the corresponding columns of $\mathbf{P}$ \via
\begin{align}    
    \mathbf{\widetilde{\Phi}} &\equiv \mathbf{P}_{:,\mathcal{C}},
    \label{eq:Phi_proj}
\end{align}
where the $\colon$\ in the subscript indicates that all rows/entries of the selected columns $\mathcal{C}$ are retained.
Since $\mathbf{\widetilde{\Phi}}$ is simply a sub-matrix of $\mathbf{P}$ (more specifically, a subset of $N_{\rm occ}$ columns of $\mathbf{P}$), $\mathbf{\widetilde{\Phi}}$ will not (in general) have orthonormal columns.
In other words, the corresponding overlap matrix,
\begin{align}    
    \mathbf{\widetilde{S}} \equiv \mathbf{\widetilde{\Phi}}^{\ast} \mathbf{\widetilde{\Phi}} , 
    \label{eq:Phi_overlap}
\end{align}
will not be equivalent to the identity matrix, \ie $\mathbf{\widetilde{S}} \neq \mathbf{I}$.
Therefore, the last step of the SCDM procedure is orthogonalization of $\mathbf{\widetilde{\Phi}}$ to obtain $\mathbf{\Phi}$, which contains the final set of orthonormal LOs ($\{{\phi}_{i}(\mathbf{r})\}_{i=1}^{N_{\rm occ}}$)) on the real-space grid.
This is accomplished using the so-called L{\"o}wdin~\cite{lowdin1950nonorthogonality} (or symmetric) orthogonalization protocol, 
\begin{align}    
    \mathbf{\Phi}=\mathbf{\widetilde{\Phi}}\mathbf{\widetilde{S}}^{-\frac{1}{2}} 
    \label{eq:Phi_ortho}
\end{align}
and can be interpreted as picking the orthonormal basis spanning the occupied space that most closely resembles the selected columns (proto-LOs).~\cite{Note2,Car57}

For many condensed-phase systems, the size of $\mathbf{P}$ ($N_{\rm grid} \times N_{\rm grid}$) prohibits its direct storage and manipulation (as the cost of the above procedure scales as $\mathcal{O}(N_{\rm grid}^3)$).
Since $\mathbf{\Psi}$ has orthonormal columns and $\mathbf{Q}$ is unitary, it can be shown~\cite{Dam15} that a rank-revealing QR factorization of the $N_{\rm occ} \times N_{\rm grid}$ $\mathbf{\Psi}^{\ast}$ matrix, \ie
\begin{align}
    \mathbf{\Psi}^{\ast}\mathbf{\Pi}=\mathbf{Q}\mathbf{R} ,
    \label{eq:qr_psi}
\end{align}
is equivalent to the SCDM procedure outlined above, and reduces the sizes of $\mathbf{Q}$ and $\mathbf{R}$ to $N_{\rm occ} \times N_{\rm occ}$ and $N_{\rm occ} \times N_{\rm grid}$, respectively.
Doing so also reduces the computational cost of the SCDM procedure from $\mathcal{O}(N_{\rm grid}^3)$ to $\mathcal{O}(N_{\rm occ}^2N_{\rm grid})$.
As before, $\mathcal{C}$ is extracted from $\mathbf{\Pi},$ the set of proto-LOs on the real-space grid are constructed as (\cf Eq.~\eqref{eq:Phi_proj})
\begin{align}
    \mathbf{\widetilde{\Phi}}=\mathbf{\Psi}\mathbf{\Psi}_{:,\mathcal{C}}^{\ast} ,
    \label{eq:phi_otc}
\end{align}
and the symmetric orthogonalization procedure in Eq.~\eqref{eq:Phi_ortho} is used to generate the final set of LOs.
%
%
\begin{algorithm}[H] 
    \caption{SCDM Procedure}
    {\footnotesize \textbf{Input:} Grid representation of COs ($\mathbf{\Psi}$)} \\
    {\footnotesize \textbf{Output:} Grid representation of LOs ($\mathbf{\Phi}$)}
    \begin{algorithmic}[1] 
        \State {\footnotesize Obtain $\mathbf{\Pi}$ from pivoted QR factorization of $\mathbf{\Psi}^{\ast}$}
        \State {\footnotesize Extract $\mathcal{C}$ from $\mathbf{\Pi}$}
        \State {\footnotesize Select proto-LOs: $\mathbf{\widetilde{\Phi}}=\mathbf{\Psi}\mathbf{\Psi}_{:,\mathcal{C}}^{\ast}$}
        \State {\footnotesize Orthogonalize proto-LOs: $\mathbf{\Phi}=\mathbf{\widetilde{\Phi}}\mathbf{\widetilde{S}}^{-\frac{1}{2}}$}
  \end{algorithmic}
  \label{alg:scdm1}
\end{algorithm}
%
%

This SCDM procedure is summarized in Algorithm~\ref{alg:scdm1}, and has been successfully applied to finite-gap condensed-phase systems in which the first Brillouin zone can be accurately sampled at the $\Gamma$-point only.~\cite{Dam15,Dam17b}
Extensions of this procedure to simulations which require $\mathbf{k}$-point sampling (\ie SCDM-k~\cite{Dam17}) as well as condensed-phase systems with entangled bands~\cite{Dam18} are described in the literature and will not be discussed here.
However, an extension of the SCDM procedure to the quantum chemical (\ie atomic-orbital based) treatment of molecular systems has not been accomplished to date, and will be the primary focus of this work.

\subsection{Extension of SCDM to Molecular Quantum Mechanics \label{sec:extension}}

We now turn our discussion to an extension of the SCDM method to the generation of localized orbitals in systems described using a non-orthogonal atomic-orbital (AO) basis set, \ie the standard approach for performing high-level quantum chemical calculations on molecules.
In doing so, we will describe the theoretical and algorithmic changes that are required to extend the procedure described above into the quantum chemical framework, and present three different variants of SCDM that can be used in conjunction with an underlying AO basis set.

Without loss of generality, we will again focus our discussion on the closed-shell (spin-unpolarized) case (\ie the restricted HF or restricted KS formalism), and only quantities that correspond to the $\alpha$-spin manifold will be considered below.
We start with a set of $N_{\rm occ}$ occupied CMOs, $\{\psi_i(\mathbf{r})\}_{i=1}^{N_{\rm occ}}$, that have been obtained \via the solution of the non-linear HF (or KS) equations.
These canonical \textit{molecular} orbitals are orthonormal (\ie $\braket{\psi_i(\mathbf{r})|\psi_j(\mathbf{r})}=\delta_{ij}$) and can be expanded in an underlying AO basis set as follows:
\begin{align}
    \ket{\psi_i} = \sum_{\mu}^{N_{\rm AO}} \ket{\chi_{\mu}} C_{\mu i} .
    \label{eq:cmo}
\end{align}
In this expression, the summation is over all $N_{\rm AO}$ atom-centered basis functions, $\{\chi_\mu(\mathbf{r})\}_{\mu=1}^{N_{\rm AO}}$, which are (in general) non-orthogonal, \ie
\begin{align}
    S_{\mu\nu} \equiv \braket{\chi_{\mu}|\chi_{\nu}} \neq \delta_{\mu\nu} .
    \label{eq:ao_overlap}
\end{align}
In analogy to the use of $\mathbf{\Psi}$ above (\ie the matrix which represented the COs in real space), we will let $\mathbf{C}$ be an $N_{\rm AO} \times N_{\rm occ}$ matrix representing the CMOs in an AO basis and $\mathbf{S}$ be the corresponding AO overlap matrix (of size $N_{\rm AO} \times N_{\rm AO}$).
Through the SCDM procedure (several variants of which are described below), we seek to obtain a set of orthogonal LMOs, $\{\phi_i(\mathbf{r})\}_{i=1}^{N_{\rm occ}}$, that can also be expanded in the underlying AO basis, \ie
\begin{align}
    \ket{\phi_i} = \sum_{\mu}^{N_{\rm AO}} \ket{\chi_{\mu}} X_{\mu i} = \sum_{\mu}^{N_{\rm AO}} \sum_{j}^{N_{\rm occ}} \ket{\chi_{\mu}} C_{\mu j} U_{ji} ,
    \label{eq:lmo}
\end{align}
in which the LMO coefficients, $X_{\mu i}$, are related to the original CMO coefficients \via a unitary (orthogonal) transformation.
In matrix form, this can be written as $\mathbf{X} = \mathbf{C}\mathbf{U}$, where $\mathbf{X}$ is an $N_{\rm AO} \times N_{\rm occ}$ matrix representing the LMOs in an AO basis and $\mathbf{U}$ is an $N_{\rm occ} \times N_{\rm occ}$ matrix that maintains orthogonality when transforming between LMOs and CMOs.

\subsubsection{The \SCDMM Procedure: SCDM in an Atomic Orbital Basis (Mulliken Variant) \label{sec:scdm-m}}

When working in a non-orthogonal AO basis set, additional care must be taken when defining the density matrix.\cite{May92,hea98}
As such, there are many possible variants of the SCDM method in an AO basis set.
Our first variant will apply the procedure described above (with appropriate modifications) to $\mathbf{\overline{P}} \equiv \mathbf{P}\mathbf{S}$; since $\mathbf{P}\mathbf{S}$ is the central quantity used in the popular Mulliken population analysis scheme,~\cite{mul55} we refer to this variant as the \SCDMM procedure.

In the \SCDMM procedure (which is summarized in Algorithm~\ref{alg:scdmm}), one starts by performing a pivoted QR factorization of $\mathbf{\overline{P}} \equiv \mathbf{P}\mathbf{S}$, \ie $\mathbf{\overline{P}}\mathbf{\Pi}=\mathbf{Q}\mathbf{R}$.
With the permutation matrix in hand (and hence $\mathcal{C}$, which is extracted from $\mathbf{\Pi}$, \textit{vide supra}), a set of non-orthogonal proto-LMOs ($\{{\widetilde{\phi}}_{i}(\mathbf{r})\}_{i=1}^{N_{\rm occ}}$) are selected \via (\cf Eq.~\eqref{eq:Phi_proj}):
\begin{align}
    \mathbf{\widetilde{X}}=\mathbf{\overline{P}}_{:,\mathcal{C}} ,
\end{align}
in which $\mathbf{\widetilde{X}}$ is an $N_{\rm AO} \times N_{\rm occ}$ matrix representing the proto-LMOs in the underlying AO basis.
In analogy to Algorithm~\ref{alg:scdm1}, the non-orthogonal set of proto-LMOs (with corresponding overlap matrix $\mathbf{\widetilde{S}}=\mathbf{\widetilde{X}}^{\ast}\mathbf{S}\mathbf{\widetilde{X}}$) are then symmetrically orthogonalized to obtain the final set of orthogonal LMOs ($\{{\phi}_{i}(\mathbf{r})\}_{i=1}^{N_{\rm occ}}$), which are represented (in the AO basis) by the $\mathbf{X}=\widetilde{\mathbf{X}}\widetilde{\mathbf{S}}^{-\frac{1}{2}}$ coefficient matrix (see Eq.~\eqref{eq:lmo}).
While formally any orthogonalization procedure could be used here to generate the final \SCDMM LMOs, we follow the protocol established in the condensed-phase variant of SCDM by employing symmetric orthogonalization as this scheme yields orthonormal LMOs that most closely resemble the proto-LMOs (see Sec.~\ref{sec:review}).
Since the \SCDMM procedure is performed in an AO basis, the associated computational cost scales as $\mathcal{O}(N^3_{\rm AO})$; because $N_{\rm AO} \ll N_{\rm grid}$, the QR decomposition can be directly applied to $\mathbf{\overline{P}}$ without the prohibitive memory/storage costs encountered in Sec.~\ref{sec:review}.~\cite{NoteCost}
%
%
\begin{algorithm}[H] 
    \caption{\SCDMM Procedure}
    {\footnotesize \textbf{Input:} CMO coefficient ($\mathbf{C}$) and AO overlap ($\mathbf{S}$) matrices} \\
    {\footnotesize \textbf{Output:} LMO coefficient matrix ($\mathbf{X}$)}
    \begin{algorithmic}[1] 
        \State {\footnotesize Construct $\mathbf{P}=\mathbf{C}\mathbf{C}^{\ast}$ and $\mathbf{\overline{P}}=\mathbf{P}\mathbf{S}$}
        \State {\footnotesize Obtain $\mathbf{\Pi}$ from pivoted QR factorization~\cite{Note3} of $\mathbf{\overline{P}}$}
        \State {\footnotesize Extract $\mathcal{C}$ from $\mathbf{\Pi}$}
        \State {\footnotesize Select proto-LMOs: $\mathbf{\widetilde{X}}=\mathbf{\overline{P}}_{:,\mathcal{C}}$}
        \State {\footnotesize Orthogonalize proto-LMOs: $\mathbf{X}=\mathbf{\widetilde{X}}\mathbf{\widetilde{S}}^{-\frac{1}{2}}$}
  \end{algorithmic}
  \label{alg:scdmm}
\end{algorithm}
%
%

Since \SCDMM selects a well-conditioned set of columns from the $\mathbf{P}\mathbf{S}$ matrix, this procedure is equivalent to selecting a well-conditioned subset of projected atomic orbitals (PAOs).
This follows from the definition of PAOs ($\{\widetilde{\chi}_\mu(\mathbf{r})\}_{\mu=1}^{N_{\rm AO}}$) as AOs that have been projected onto the occupied space, \ie
\begin{align}
    \ket{\widetilde{\chi}_{\lambda}} &=\sum_{i}^{N_{\rm occ}} \ket{\psi_i} \braket{\psi_i|\chi_{\lambda}} = \sum_{i}^{N_{\rm occ}} \sum_{\mu\nu}^{N_{\rm AO}} \ket{\chi_{\mu}} C_{\mu i}C_{\nu i}^{\ast}\braket{\chi_{\nu}|\chi_{\lambda}} \nonumber \\
    &= \sum_{\mu\nu}^{N_{\rm AO}} \ket{\chi_{\mu}} P_{\mu\nu} S_{\nu\lambda} = \sum_{\mu}^{N_{\rm AO}} \ket{\chi_{\mu}} (\mathbf{P}\mathbf{S})_{\mu\lambda} ,
    \label{eq:pao}
\end{align}
where we have used Eqs.~\eqref{eq:cmo}--\eqref{eq:ao_overlap} and the definition $\mathbf{P}\equiv\mathbf{C}\mathbf{C}^{\ast}$.
As such, the PAOs correspond to a redundant set of $N_{\rm AO}$ functions (with rank equal to $N_{\rm occ}$) that formally span the occupied space and whose coefficient matrix is simply $\mathbf{P}\mathbf{S}$.
Since PAOs inherit a certain degree of locality from the underlying AO basis, these projected orbitals have been proposed as redundant (non-orthogonal) representations for the occupied and virtual subspaces in local electron correlation methods~\cite{sae88,Sae93,Mas98,Sch99,lee00b,wer03,dis05,ham96,SCH00,Sch01,sch02}.
In the virtual case, one projects \textit{out} the occupied space and is left with a set of local virtuals that is only slightly redundant in the complete basis set (CBS) limit (\ie where the number of virtual orbitals $N_{\rm virt} \simeq N_{\rm AO}$).
Since $N_{\rm AO} \gg N_{\rm occ}$ in the CBS limit, a PAO representation of the occupied space (see Eq.~\eqref{eq:pao}) is highly redundant; as such, PAOs have not found much use as localized occupied orbitals in local electron correlation methods (especially with the availability of orthogonal LMOs such as Boys, PM, ER, and others).
However, the \SCDMM procedure described above provides a convenient avenue towards removing this large degree of redundancy, yielding an orthogonal set of PAO-based LMOs that exactly span the occupied space (and could potentially be used in local WFT~\cite{Sae87,sae88,Sae93,ham96,Mas98,rau98,Sch99,SCH00,lee00b,Sch01,sch02,wer03,dis05,Kat08,nee09,nee09b,hel11,rip13} and DFT~\cite{Sol02,Wu09,gyg09,chen_x-ray_2010,hin11,gyg13,swartz_ab_2013,dis14,Mus15,daw15,hes17,hsi19a} methods).
%
%
\begin{figure}[t]
    \centering
    \includegraphics[width=1.0\linewidth]{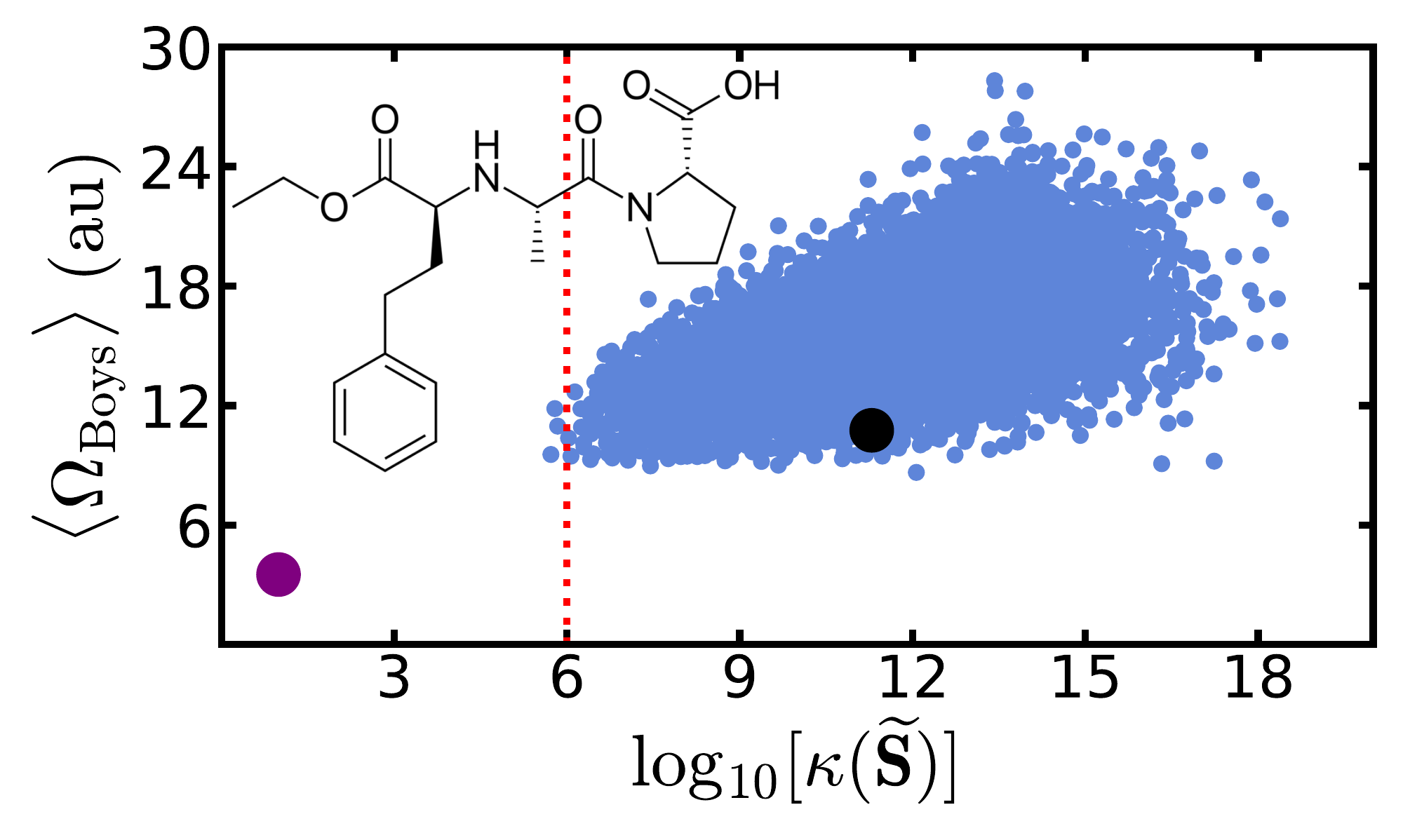}
    \caption{
    Semi-logarithmic plot of orbital locality versus condition number for $10,000$ randomly selected sets of $N_{\rm occ}$ PAOs (\ie columns of $\mathbf{P}\mathbf{S}$) for the drug molecule enalapril (see inset) at the HF/cc-pVTZ level of theory (see Sec.~\ref{sec:comp} for computational details).
    In this plot, orbital locality was measured by the average orbital variance (or second central moment, see Eq.~\eqref{eq:boys}) associated with each set of symmetrically orthogonalized PAOs, \ie $\braket{\Omega_{\rm Boys}} = (1/N_{\rm occ}) \sum_i^{N_{\rm occ}} \sigma_i^2$. 
    The condition number ($\kappa$) of the corresponding PAO overlap matrix, $\widetilde{S}_{\mu\nu} = \braket{\widetilde{\chi}_{\mu}|\widetilde{\chi}_{\nu}}$, was used as a measure of the linear dependence in these randomly chosen subsets of PAOs (with the dashed vertical red line denoting a common numerical threshold used to indicate linear dependence in the underlying AO basis).
    Also included are the results of the \SCDMM procedure (filled purple circle, Algorithm~\ref{alg:scdmm}) and a deterministic procedure which selects PAOs with the largest norms (filled black circle, see text for more details).
    For clarity, these points have been made larger in the plot.
    }
    \label{fig:rand}
\end{figure}
%
%

Here, we emphasize that not all subsets of $N_{\rm occ}$ columns of $\mathbf{P}\mathbf{S}$ (\ie $N_{\rm occ}$ PAOs) will form a well-conditioned set; hence, most of these subsets will not result in an orthonormal set of localized orbitals.
To illustrate this point, we selected $10,000$ subsets of $N_{\rm occ}$ PAOs uniformly at random and computed their condition number as well as the locality of the resulting orthonormal set of orbitals.
As depicted in Fig.~\ref{fig:rand} for the drug molecule enalapril (\ce{C20H28N2O5}, chosen for its wide functional group diversity), almost all of the randomly selected sets ($99.97\%$) had condition numbers (\ie $\kappa(\mathbf{\widetilde{S}})$ of the corresponding PAO overlap matrix) that were larger than $\kappa(\mathbf{S}) = 10^{6}$, a common numerical threshold used to indicate linear dependence in the underlying AO basis.
In the best case scenario, the condition number of the randomly chosen PAO subsets (\ie $\kappa(\mathbf{\widetilde{S}}) \approx 10^{6}$) was more than five orders of magnitude larger than that of the well-conditioned subset of PAOs chosen by the \SCDMM procedure (filled purple circle in Fig.~\ref{fig:rand}).
After symmetric orthogonalization of the randomly chosen proto-LMO sets, we find that every set of LMOs is significantly more delocalized (by at least $2\times$ in the $\braket{\Omega_{\rm Boys}}$ metric) than the corresponding set of \SCDMM generated LMOs.
This follows from the fact that the (relatively ill-conditioned) random sets contain strongly overlapping PAOs which will acquire extensive tails during the orthogonalization procedure; in contrast, the well-conditioned set of PAOs chosen by the \SCDMM procedure is minimally perturbed by the symmetric orthogonalization, and the underlying locality of the PAOs is reflected in the final set of LMOs.

Although the \SCDMM procedure cannot guarantee selection of the \textit{most} well-conditioned set of proto-LMOs (as this is a provably hard problem~\cite{Civ09}), the heuristics of the column-pivoted QR algorithm nevertheless lead to a set of well-localized orbitals.
In fact, it is this column-pivoted QR factorization that is responsible for the efficient generation of a well-conditioned set of proto-LMOs---the crucial first step in the generation of a well-localized and orthogonal set of LMOs that exactly span the occupied space.
To see this more clearly, it is helpful to understand how the column-pivoted QR factorization employed in the \SCDMM procedure actually guides the selection of PAOs as proto-LMOs.
In this case, the algorithm selects the column of $\mathbf{P}\mathbf{S}$ corresponding to the PAO with the largest norm (\textit{i.e.,} the PAO with the most significant overlap with the occupied space), projects this PAO out of $\mathbf{P}\mathbf{S}$, and then repeats these two steps until completion.
In doing so, it is this intermediary projection step between selecting columns that is the key ingredient when determining a well-conditioned set of PAOs.
To illustrate this point, consider a simple deterministic algorithm (not invoking a column-pivoted QR factorization) in which the columns of $\mathbf{P}\mathbf{S}$ corresponding to the $N_{\rm occ}$ PAOs with the largest norms, \ie the PAOs with the most significant overlap with the occupied space, are selected as proto-LMOs.
As shown in Fig.~\ref{fig:rand}, the resulting set of proto-LMOs using this procedure are ill-conditioned (filled black circle, $\kappa(\mathbf{\widetilde{S}}) \approx 10^{11}$), and therefore cannot be reliably orthogonalized to generate a final set of LMOs.
As such, it is clear that column-pivoted QR factorization (and hence the \SCDMM procedure) outperforms na\"ive/random as well as simple deterministic selection methods, and is therefore a powerful technique for removing redundancies and generating an orthogonal set of LMOs that still retains the local character of the PAOs.

\subsubsection{The \SCDML Procedure: SCDM in an Atomic Orbital Basis (L{\"o}wdin Variant) \label{sec:scdm-l}}

When working in a non-orthogonal AO basis set, it is often useful to (symmetrically) orthogonalize the basis \via  
\begin{align}
    \ket{\chi_{\nu}'}=\sum_{\mu}^{N_{\rm AO}}\ket{\chi_{\mu}} S^{-\frac{1}{2}}_{\mu\nu} ,
    \label{eq:oao}
\end{align}
in which $\{\chi_{\nu}'(\mathbf{r})\}_{\nu=1}^{N_{\rm AO}}$ is the set of orthogonalized atomic orbitals (OAOs).
In this basis, $\mathbf{S}^{\frac{1}{2}}\mathbf{P}\mathbf{S}^{\frac{1}{2}}$ is the corresponding (one-particle) density matrix.
Hence, our second variant of SCDM will apply the procedure described above (again with appropriate modifications) to $\mathbf{\overline{P}} \equiv \mathbf{S}^{\frac{1}{2}}\mathbf{P}\mathbf{S}^{\frac{1}{2}}$; since $\mathbf{S}^{\frac{1}{2}}\mathbf{P}\mathbf{S}^{\frac{1}{2}}$ is the central quantity used in L{\"o}wdin population analysis,~\cite{lowdin1950nonorthogonality} we will refer to this variant as the \SCDML procedure.

%
%
\begin{algorithm}[H] 
    \caption{\SCDML Procedure}
    {\footnotesize \textbf{Input:} CMO coefficient ($\mathbf{C}$) and AO overlap ($\mathbf{S}$) matrices} \\
    {\footnotesize \textbf{Output:} LMO coefficient matrix ($\mathbf{X}$)}
    \begin{algorithmic}[1] 
        \State {\footnotesize Construct $\mathbf{P}=\mathbf{C}\mathbf{C}^{\ast}$ and $\mathbf{\overline{P}}=\mathbf{S}^{\frac{1}{2}}\mathbf{P}\mathbf{S}^{\frac{1}{2}}$}
        \State {\footnotesize Obtain $\mathbf{\Pi}$ from pivoted QR factorization of $\mathbf{\overline{P}}$}
        \State {\footnotesize Extract $\mathcal{C}$ from $\mathbf{\Pi}$}
        \State {\footnotesize Select proto-LMOs: $\mathbf{\widetilde{X}}=\mathbf{\overline{P}}_{:,\mathcal{C}}$}
        \State {\footnotesize Orthogonalize/back transform proto-LMOs: $\mathbf{X}\!=\!\mathbf{S}^{-\frac{1}{2}}\mathbf{\widetilde{X}}\mathbf{\widetilde{S}}^{-\frac{1}{2}}$}
  \end{algorithmic}
  \label{alg:scdml}
\end{algorithm}
%
%
As illustrated in Algorithm~\ref{alg:scdml}, the \SCDML procedure (with a computational cost that scales as $\mathcal{O}(N^3_{\rm AO})$) is similar to the \SCDMM procedure in Algorithm~\ref{alg:scdmm}; the only differences are that $\mathbf{\overline{P}} \equiv \mathbf{S}^{\frac{1}{2}}\mathbf{P}\mathbf{S}^{\frac{1}{2}}$ and the \SCDML procedure requires an additional back-transformation step (from OAOs to AOs \via Eq.~\eqref{eq:oao}) for consistency with Eq.~\eqref{eq:lmo}.
Since \SCDML selects a well-conditioned set of columns from $\mathbf{S}^{\frac{1}{2}}\mathbf{P}\mathbf{S}^{\frac{1}{2}}$, this procedure is equivalent to selecting a well-conditioned subset of projected (symmetrically-) orthogonalized atomic orbitals (POAOs).
In analogy to the \SCDMM procedure, this follows from the definition of POAOs ($\{\widetilde{\chi}'_\mu(\mathbf{r})\}_{\mu=1}^{N_{\rm AO}}$) as OAOs that have been projected onto the occupied space, \ie
\begin{align}
    \ket{\widetilde{\chi}_{\lambda}'} &=\sum_{i}^{N_{\rm occ}} \ket{\psi_i} \braket{\psi_i|\chi_{\lambda}'} = \sum_{i}^{N_{\rm occ}} \sum_{\mu\nu}^{N_{\rm AO}} \ket{\chi_{\mu}} C_{\mu i}C_{\nu i}^{\ast}\braket{\chi_{\nu}|\chi_{\lambda}'} \nonumber \\
    &= \sum_{\mu\nu\sigma\xi}^{N_{\rm AO}} \ket{\chi_{\sigma}'} S^{\frac{1}{2}}_{\sigma\mu}P_{\mu\nu} S^{\frac{1}{2}}_{\nu\xi}\braket{\chi_{\xi}'|\chi_{\lambda}'} = \sum_{\mu\nu\sigma}^{N_{\rm AO}} \ket{\chi_{\sigma}'} S^{\frac{1}{2}}_{\sigma\mu}P_{\mu\nu} S^{\frac{1}{2}}_{\nu\lambda} \nonumber \\
    &= \sum_{\sigma}^{N_{\rm AO}} \ket{\chi_{\sigma}'} (\mathbf{S}^{\frac{1}{2}}\mathbf{P}\mathbf{S}^{\frac{1}{2}})_{\sigma\lambda} ,
    \label{eq:poao}
\end{align}
where we have used the inverse of Eq.~\eqref{eq:oao} and the fact that $S_{\xi\lambda}' = \braket{\chi_\xi' | \chi_\lambda'} = \delta_{\xi\lambda}$ for the set of OAOs. 
LMOs generated with the \SCDMM and \SCDML procedures will be compared and contrasted below in Secs.~\ref{sec:prop}--\ref{sec:interpret}.

\subsubsection{The \SCDMG Procedure: SCDM in an Atomic Orbital Basis (Grid Variant) \label{sec:scdm-g}}

While working directly with the density matrix in an AO basis may be conceptually straightforward, the \SCDMM and \SCDML variants described above represent a fundamental restriction of the SCDM procedure.
In particular, the proto-LMOs selected using these variants are limited to subsets of PAOs (\SCDMM) or POAOs (\SCDML), while the original condensed-phase SCDM procedure (see Sec.~\ref{sec:review}) has significantly more flexibility in its choice of proto-LMOs.
In this regard, one interpretation of Eq.~\eqref{eq:Phi_proj} (or Eq.~\eqref{eq:phi_otc}) is that SCDM constructs a set of proto-LMOs by carefully selecting a subset of Dirac $\delta$ functions projected onto the occupied space. 
To take advantage of this increased flexibility (and therefore better adapt to the underlying physics described by a real-space representation of the density matrix), we now develop a grid-based variant of SCDM within an AO framework, which will henceforth be referred to as \SCDMG.
When working in real space, we remind the reader that the pivoted QR factorization step in the SCDM approach can equivalently~\cite{Dam15} be applied to the $\mathbf{P}$ and $\mathbf{\Psi}^{\ast}$ matrices (\cf Eqs.~\eqref{eq:qr} and \eqref{eq:qr_psi}); since the latter scales more favorably, we will only apply the \SCDMG procedure to $\mathbf{\Psi}^{\ast}$ below.

To begin, we represent each AO basis function on a set of real-space grid points ($\{\mathbf{r}_p\}_{p=1}^{N_{\rm grid}}$) as follows:
\begin{align}
    \ket{\chi_\mu} = \sum_p^{N_{\rm grid}} \ket{\mathbf{r}_p}  W_{p\mu} ,
    \label{eq:togrid}
\end{align}
in which $W_{p\mu}$ is the numerical value of the $\mu$-th basis function on the $p$-th grid point.
As illustrated in Algorithm~\ref{alg:scdmg}, the first step of the \SCDMG procedure is to use Eq.~\eqref{eq:togrid} to represent the CMOs on the real-space grid, \ie $\mathbf{\Psi}=\mathbf{W}\mathbf{C}$.
The \SCDMG procedure now follows Steps 1-3 of Algorithm~\ref{alg:scdm1} by performing a pivoted QR factorization of $\mathbf{\Psi}^{\ast}$ followed by proto-LMO selection/construction \via $\mathbf{\widetilde{\Phi}}=\mathbf{\Psi}\mathbf{\Psi}_{:,\mathcal{C}}^{\ast}$ (see Eqs.~\eqref{eq:qr_psi}--\eqref{eq:phi_otc}).
The \SCDMG procedure then continues by re-expressing these proto-LMOs in the underlying AO basis by applying the following transformation:
\begin{align}
    \mathbf{\widetilde{X}} = \mathbf{S}^{-1} \mathbf{W}^T \mathbf{\widetilde{\Phi}} .
    \label{eq:toao}
\end{align}
This is followed by symmetric orthogonalization to yield the final LMO coefficient matrix, $\mathbf{X}=\mathbf{\widetilde{X}}\mathbf{\widetilde{S}}^{-\frac{1}{2}}$, in accordance with Eq.~\eqref{eq:lmo} and the procedure established in Sec.~\ref{sec:review}. 
Notably, by construction the proto-LMOs, and therefore the LMOs, produced by the \SCDMG method are invariant to any changes that preserve $\mathbf{\Psi}$.
%
%
\begin{algorithm}[H] 
    \caption{\SCDMG Procedure}
    {\footnotesize \textbf{Input:} CMO coefficient ($\mathbf{C}$) and AO overlap ($\mathbf{S}$) matrices} \\
    {\footnotesize \textbf{Output:} LMO coefficient matrix ($\mathbf{X}$)}
    \begin{algorithmic}[1] 
        \State {\footnotesize Transform CMOs to grid representation: $\mathbf{\Psi}=\mathbf{WC}$}
        \State {\footnotesize Obtain $\mathbf{\Pi}$ from pivoted QR factorization of $\mathbf{\Psi}^{\ast}$}
        \State {\footnotesize Extract $\mathcal{C}$ from $\mathbf{\Pi}$}
        \State {\footnotesize Select proto-LMOs: $\mathbf{\widetilde{\Phi}}=\mathbf{\Psi}\mathbf{\Psi}^{\ast}_{:,\mathcal{C}}$}
        \State {\footnotesize Transform proto-LMOs to AO basis: $\mathbf{\widetilde{X}}=\mathbf{S^{-1}W}^{T}\mathbf{\widetilde{\Phi}}$}
        \State {\footnotesize Orthogonalize proto-LMOs: $\mathbf{X}=\mathbf{\widetilde{X}}\mathbf{\widetilde{S}}^{-\frac{1}{2}}$}
  \end{algorithmic}
  \label{alg:scdmg}
\end{algorithm}
%
%

The most time consuming step of the \SCDMG procedure depicted in Algorithm~\ref{alg:scdmg}, is the pivoted QR factorization, which has an associated computational cost that scales as $\mathcal{O}(N_{\rm occ}^2 N_{\rm grid})$~\cite{Note4}; as such, \SCDMG will typically be more expensive than \SCDMM and \SCDML (both of which scale as $\mathcal{O}(N_{\rm AO}^3)$).
Unlike the original condensed-phase version of SCDM described above (which also scales as $\mathcal{O}(N_{\rm occ}^2 N_{\rm grid})$), the real-space grid employed in the \SCDMG procedure can be completely decoupled from the underlying energy calculation, and can therefore be chosen to balance computational cost versus flexibility in the selection of proto-LMOs.
For instance, very few (if any) grid points need to be placed in regions of space where the electron density is relatively small, as these grid points are unlikely to be selected by the \SCDMG procedure.
In the same breath, it is very important to have grid points located in regions of significant electron density (\ie the support of the occupied space), as these points are likely to be selected when generating proto-LMOs.
As such, standard DFT grid implementations~\cite{gil93,tre95,leb1999} used for numerical quadrature in molecular quantum mechanics packages are an attractive option, and can easily be re-purposed for LMO generation using the \SCDMG procedure.
In addition, several techniques which have been devised to reduce the cost of the condensed-phase variant of SCDM (\eg non-uniform random sub-sampling of the grid prior to performing the column-pivoted QR factorization~\cite{Dam17b}) could also be used here to increase computational efficiency. 
For an analysis of the grid dependence of the \SCDMG procedure, see Sec.~\ref
{sec:grid_appendix}.

While there are fundamental differences between the \SCDMM, \SCDML, and \SCDMG procedures, all three inherit the intrinsic benefits of the original condensed-phase SCDM approach.
In particular, all three of these SCDM variants are direct (non-iterative) methods with a fixed computational cost and do not require an initial guess.
Although this family of SCDM variants generates a set of LMOs that do not correspond to a solution of an optimization problem of a chosen metric (\eg $\Omega_{\rm Boys}$, $\Omega_{\rm PM}$, etc.), each procedure implicitly infers information about the molecular system \via the density matrix during LMO construction.
As we will demonstrate in the following section, these SCDM-generated LMOs are quite similar to other popular orbital localization schemes (\eg Boys, PM, etc.) when dealing with a wide range of molecular systems.

\section{Results and Discussion  \label{sec:rd}}

\subsection{Computational Details  \label{sec:comp}}

The \SCDMM, \SCDML, and \SCDMG algorithms have been implemented in the \texttt{PySCF} electronic structure package~\cite{sun18}, using an interface to the column-pivoted QR routine \texttt{DGEQP3} in LAPACK.~\cite{and99}
All calculations were performed at the mean-field HF level in conjunction with the correlation-consistent cc-pVXZ and aug-cc-pVXZ (with X = D, T, Q) AO basis sets of Dunning and co-workers;~\cite{dun89,ken92} all self-consistent field (SCF) calculations used the default convergence criteria in \texttt{PySCF}.
Unless otherwise specified, cc-pVTZ was employed as the default AO basis set throughout the remainder of this work.
Applications to standard semi-local DFT functionals (as opposed to HF) would not require any modifications to the code or theoretical approach described above.
Molecular geometries for enalapril and the hydrocarbon (alkanes and alkenes) test molecules were optimized at the HF/cc-pVTZ level, while those for the glycine-based polypeptides were generated using the \texttt{PeptideBuilder} package.~\cite{tie13} 
Unless otherwise specified, we employ the standard frozen core (FC) approximation throughout, in which the [1s] core orbitals (for all first-row (CNOF) elements) are not considered during the \textit{a posteriori} processing of the HF orbitals, \ie the density matrices utilized herein will be constructed from the \textit{active} (or valence) occupied orbitals only. 
This was done as it is often desirable to separate the core and valence spaces during post-HF electron correlation methods (in which the FC approximation is almost always employed); inclusion of the core orbitals would not require any modifications to the code or theoretical approach described above.
The real-space grids utilized in the \SCDMG procedure were constructed according to the radial grid generation scheme of Treutler and Ahlrichs~\cite{tre95} (with $200$ radial grid points per atom) and the angular grid generation scheme of Lebedev~\cite{leb1999} (with $1,454$ angular grid points per radial shell).
All grids were pruned according to the scheme implemented in the \texttt{NWChem} package.~\cite{nwchem} As seen in Fig.~\ref{fig:grid1}, the locality of the LMOs produced by the \SCDMG procedure is rather insensitive to the size of the underlying real space grid once it is sufficiently dense (\ie a medium-quality grid containing $(50,302)$ [H] and $(75,302)$ [C] radial and angular grid points before pruning). 
When generating LMOs using the Boys and PM schemes, the resulting LMOs were converged to $10^{-10}$ in the corresponding cost function (and $10^{-6}$ in the respective gradient).

\subsection{Properties of SCDM Orbitals in Molecular Systems  \label{sec:prop}}

%
%
\begin{figure}[t]
    \centering
    \includegraphics[width=0.99\columnwidth]{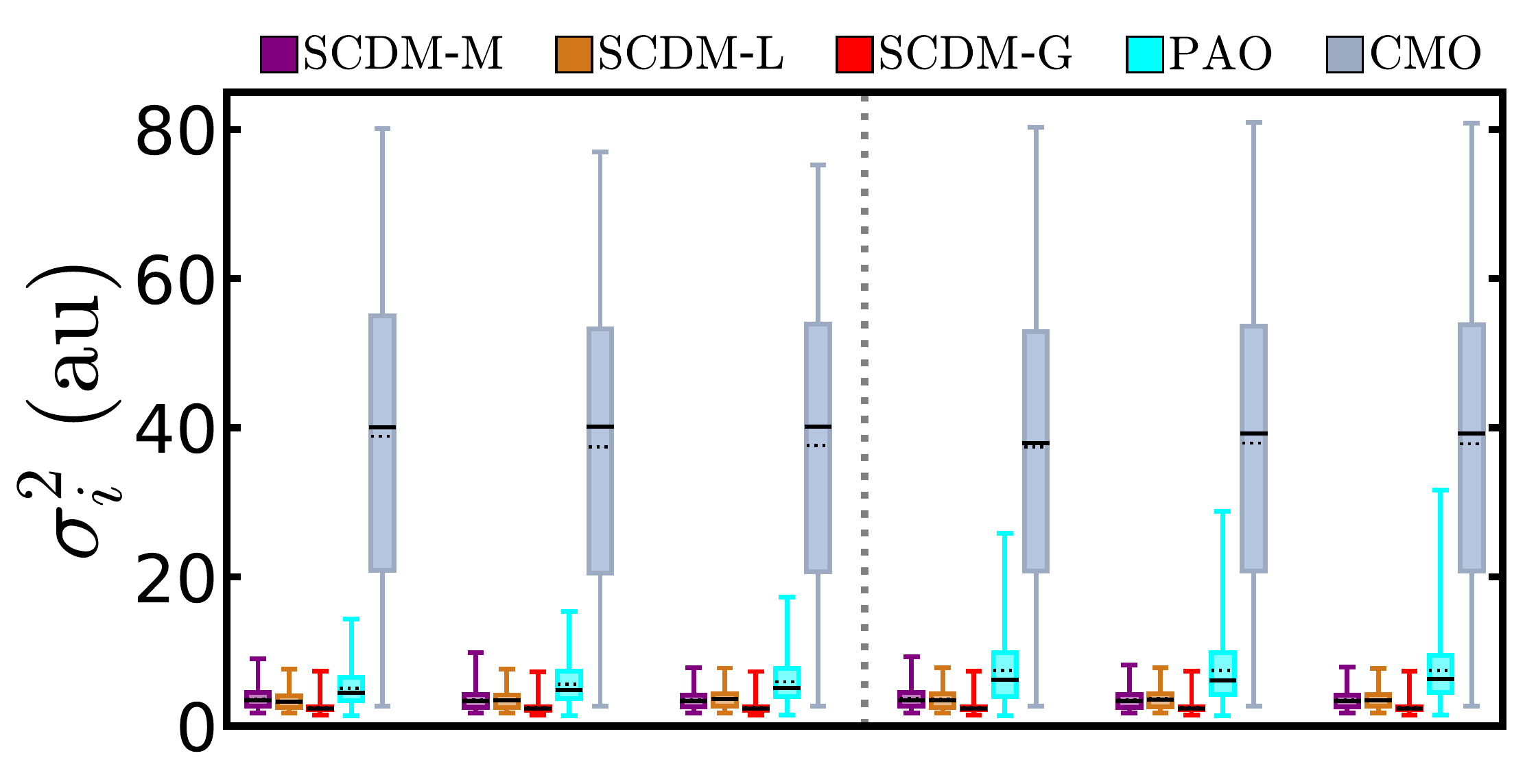} \\
    \includegraphics[width=0.99\columnwidth]{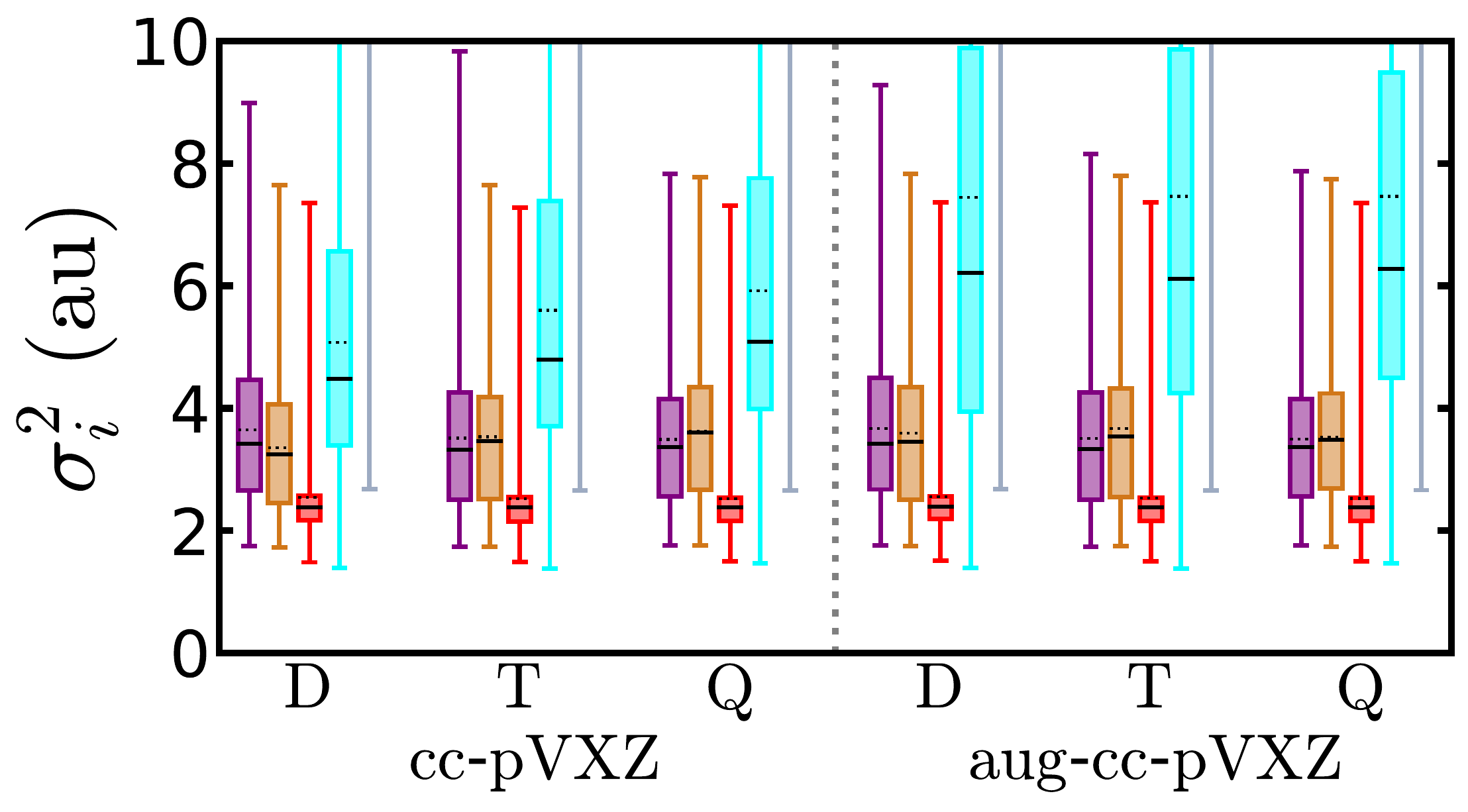}
    \caption{
    \textit{Top panel}: Box-and-whisker plots of orbital locality versus AO basis set for the three SCDM variants (\SCDMM, \SCDML, and \SCDMG), the non-orthogonal set of PAOs, and the CMOs of the drug molecule enalapril (\ce{C20H28N2O5}, see Fig.~\ref{fig:rand} and text).
    In this plot, orbital locality was measured by the orbital variance (or second central moment, see Eq.~\eqref{eq:boys}) associated with each set of orbitals, and the vertical dotted line separates the standard (cc-pVXZ with X = D, T, Q) and augmented (aug-cc-pVXZ with X = D, T, Q) Dunning-style AO basis sets.
    In each case, the whiskers extend to the corresponding minimum and maximum values, while the box demarcates the $Q_1\mathrm{-}Q_3$ ($25\%\mathrm{-}75\%$) interquartile range; the solid (dashed) lines within each box mark the position of the median (mean).
    \textit{Bottom panel}: Zoomed-in version of the top panel which focuses on the three SCDM variants and illustrates the improved locality obtained using the \SCDMG procedure.
    }
    \label{fig:basis}
\end{figure}
%
%
To investigate the performance of the SCDM variants introduced in Sec.~\ref{sec:theory} for molecular systems, we generated LMOs using the \SCDMM, \SCDML, and \SCDMG procedures (as well as the Boys and PM procedures) for a number of test systems of varying size, dimensionality, and saturation level.
To begin, we consider how the locality of the SCDM-generated LMOs depends on the underlying AO basis set.
We again use the drug molecule enalapril due to its three-dimensional structure and rich functional group diversity. 
As seen in Fig.~\ref{fig:basis}, we first observe that the SCDM-generated LMOs (all three variants) and the non-orthogonal set of PAOs are all significantly more local than the CMOs, which is to be expected since the CMOs are (by definition) delocalized across the entire enalapril molecule.
While the PAOs inherit a certain degree of locality from the underlying AO basis, we also see that the SCDM-generated LMOs are significantly more local than this redundant and non-orthogonal representation of the occupied space.
This is interesting when one considers that the LMOs generated by the \SCDMM procedure (in particular) were obtained by symmetrically orthogonalizing a subset of PAOs (see Eq.~\eqref{eq:pao} and surrounding discussion).
Here, we would stress that the \SCDMM procedure is clearly not selecting the most diffuse PAOs (as they weakly overlap with the occupied space); instead, this algorithm generates a well-conditioned set of PAOs by making a compromise between selecting PAOs that overlap most strongly with the occupied space (\ie PAOs with the largest norms) and PAOs that are distinct enough to avoid linear dependency issues.
In this regard, we would argue that this selection procedure is fairly well-balanced, as the final set of LMOs (which are generated \via a symmetric orthogonalization step that invariably leads to further delocalization) still remain more local than the non-orthogonal set of PAOs.

A quick glance at Fig.~\ref{fig:basis} also reveals that the grid-based \SCDMG procedure yields LMOs with improved overall locality over the \SCDMM and \SCDML procedures.
In particular, \SCDMG LMOs show reduced minimum and maximum $\sigma^2$ values, as well as significantly lower and more compressed interquartile ranges for every AO basis set considered in this work.
In brief, this improved locality can be attributed to the added flexibility gained by evaluating the density matrix on a real-space grid during the \SCDMG procedure, and this point will be investigated in more detail throughout the remainder of the manuscript.
Fig.~\ref{fig:basis} also demonstrates that the locality of the CMOs is largely independent of the underlying AO basis set since the occupied space in HF (like DFT) has a weak dependence on the basis and tends to converge with triple-$\zeta$ quality basis sets (\eg cc-pVTZ). 
Here, we note in passing that this is in stark contrast to the PAOs, since the dimensionality (and inherent locality) of this redundant set is directly tied to that of the underlying AO basis set. 
As expected for a robust localization procedure that generates an orthogonal set of functions which exactly spans the occupied space, the locality of the SCDM-generated LMOs (like the CMOs) remains stable with respect to systematic changes in the AO basis set.
As such, we will continue to work in the cc-pVTZ basis throughout the remainder of this work.

Since the \SCDMM (\SCDML) procedures directly rely on the PAOs (POAOs) through the choice of proto-LMOs (see Eqs.~\eqref{eq:pao} and \eqref{eq:poao}), the final \SCDMM (\SCDML) LMOs will implicitly inherit a dependence on the basis set through these projected orbitals.
As such, the locality and character of the \SCDMM (\SCDML) LMOs are not necessarily invariant with respect to certain types of changes in the basis set. 
For example, different proto-LMOs (and hence LMOs) may be produced when the basis functions are rotated with respect to the global coordinate system (\ie by rotating the molecule) or when different basis set families (\eg those with different contraction schemes) are used.
Importantly, the dependence on molecular orientation can be avoided by rotating the molecule into the canonical/standard nuclear orientation~\cite{gil93} at the very beginning of the calculation---a procedure automatically performed by many software packages.
Nevertheless, these are important theoretical (and potentially practical) issues that warrant further investigation; see Secs.~\ref{sec:basis_appendix} and~\ref{sec:orientation_appendix} for a more detailed discussion.

%
%
\begin{figure*}[th!]
    \centering
    \includegraphics[width=0.75\linewidth]{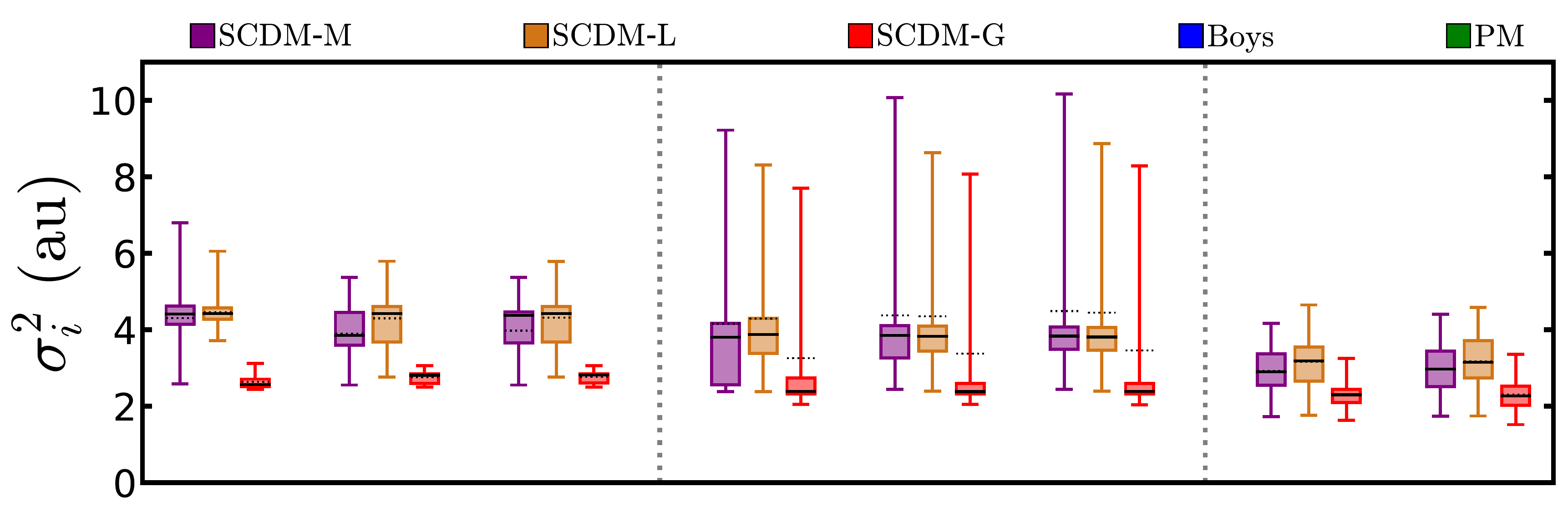} \\
    \includegraphics[width=0.75\linewidth]{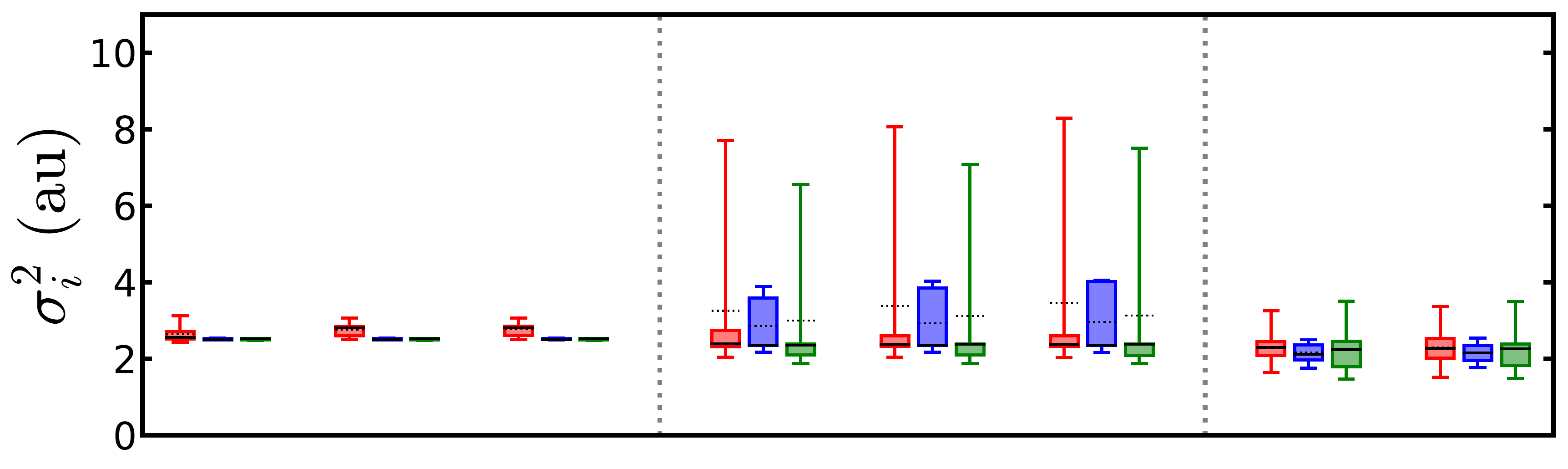} \\
    \hspace{-.16cm}
    \includegraphics[width=0.75\linewidth]{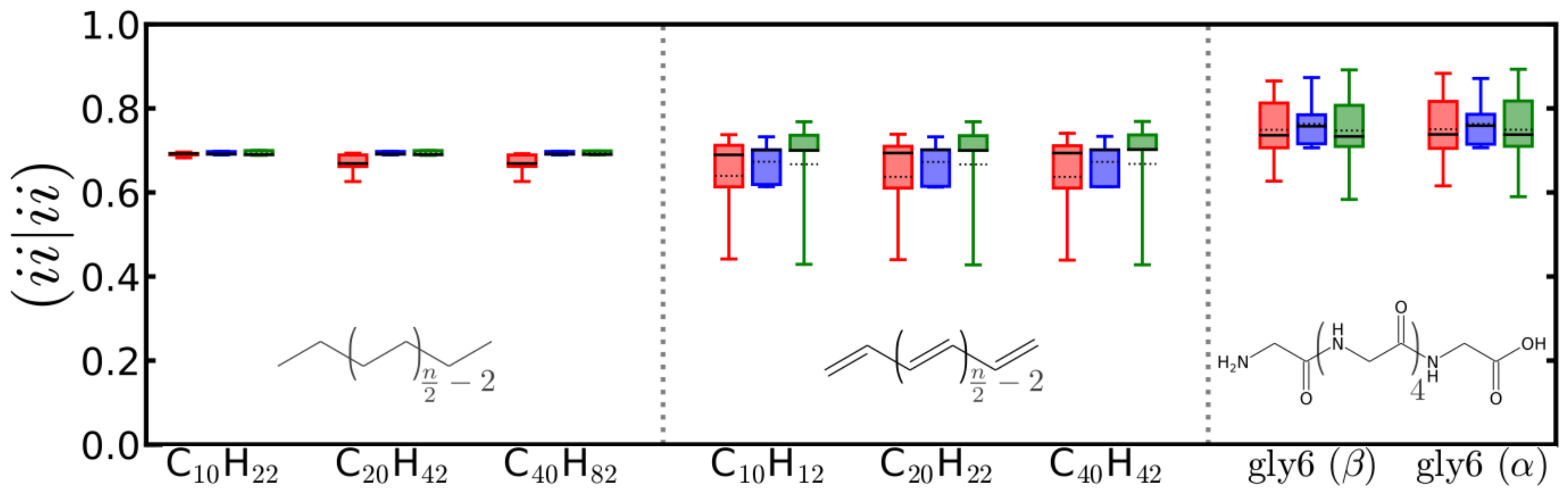}
    \caption{
    Box-and-whisker plots of orbital locality versus system type for the following test molecules of varying size, dimensionality, and saturation level (see insets): \textit{n}-alkanes (\ce{C10H22}, \ce{C20H42}, \ce{C40H82}), s-\textit{trans} alkenes (\ce{C10H12}, \ce{C20H22}, \ce{C40H42}), and hexaglycine polypeptides (gly6 ($\beta$-strand), gly6 ($\alpha$-helix)).
    \textit{Top panel}: Comparison of orbital locality (measured by the orbital variance or second central moment, see Eq.~\eqref{eq:boys}) for each test molecule using all three SCDM variants (\SCDMM, \SCDML, \SCDMG) at the HF/cc-pVTZ level of theory. 
    \textit{Middle panel}: Comparison of orbital locality (measured again by the orbital variance) for each test molecule using the \SCDMG, Boys, and PM (using Mulliken populations) methods at the HF/cc-pVTZ level of theory.
    \textit{Bottom panel}: Comparison of orbital locality (measured by the self-Coulombic repulsion of the LMOs, see Eq.~\eqref{eq:er2}) for each test molecule using the \SCDMG, Boys, and PM (using Mulliken populations) methods at the HF/cc-pVTZ level of theory.
    In each case, the whiskers extend to the corresponding minimum and maximum values of the corresponding locality metric, while the box demarcates the $Q_1\mathrm{-}Q_3$ ($25\%\mathrm{-}75\%$) interquartile range; the solid (dashed) lines within each box mark the position of the median (mean).
    }
    \label{fig:ane_ene}
\end{figure*}
%
%

As a second test, we now explore the performance of the SCDM procedures by considering the following test molecules of varying system size, dimensionality, and saturation level: \textit{n}-alkanes (\ce{C10H22}, \ce{C20H42}, \ce{C40H82}), s-\textit{trans} alkenes (\ce{C10H12}, \ce{C20H22}, \ce{C40H42}), and hexaglycine polypeptides (gly6 ($\beta$-strand), gly6 ($\alpha$-helix)).
In the top panel of Fig.~\ref{fig:ane_ene}, we use box-and-whisker plots to report the orbital variances (or second central moments, see Eq.~\eqref{eq:boys}) of the LMOs generated by the \SCDMM, \SCDML, and \SCDMG schemes for these test molecules. 
In all cases, we find that the three SCDM variants produce a set of localized orbitals that are rather insensitive to the system size; the only slight exception is the small change observed between the \ce{C10H_{\rm n}} and \ce{C20H_{\rm n}} hydrocarbon systems where asymptotic behavior seems to be reached.
Once again, we find that \SCDMG generates LMOs with enhanced bulk locality (\ie lower and more compressed interquartile ranges) and better extrema (\ie reduced minimum and maximum $\sigma^2$ values) compared to the \SCDMM and \SCDML procedures, and this trend holds across all system types considered herein.
We again attribute this improved locality to the additional flexibility associated with evaluating the density matrix on a real-space grid during the \SCDMG procedure.

In fact, the locality of the \SCDMG-generated LMOs for the alkane and polypeptide systems are comparable across all distributional metrics (\eg mean, median, interquartile range, etc.) to the locality of the LMOs generated by the iterative Boys and PM schemes.
Furthermore, this observation holds when considering either a statistical measure of locality (\ie orbital variances, see middle panel of Fig.~\ref{fig:ane_ene}) as well as an energetic measure of locality (\ie orbital self-Coulombic repulsion, see bottom panel of Fig.~\ref{fig:ane_ene}).
In this regard, it is quite encouraging that the non-iterative \SCDMG procedure---which is not based on the optimization of a specific locality metric---can produce LMOs of comparable quality to the iterative Boys scheme (which specifically minimizes the orbital variance based metric in Eq.~\eqref{eq:boys}) and the iterative PM scheme (which maximizes the population based metric in Eq.~\eqref{eq:pm}).
Although not shown, the same conclusions hold for the iterative ER scheme, which maximizes the orbital self-Coulombic repulsion measure of locality (see Eqs.~\eqref{eq:er}--\eqref{eq:er2}).

When considering the s-\textit{trans} alkenes, we again find that the LMOs produced by the \SCDMG, Boys, and PM schemes are comparable, although a comparative analysis of these orbitals is more subtle and warrants further discussion.
Here, one could argue that the electronic structure of such conjugated $\pi$-systems is significantly more delocalized, and will therefore pose additional challenges for orbital localization schemes that are not necessarily present in the extended \textit{n}-alkane and polypeptide systems.~\cite{Hoy14,Hoy16} 
When considering the middle panel of Fig.~\ref{fig:ane_ene}, one sees that the \SCDMG and PM schemes produce rather similar LMOs, with enhanced bulk locality (\ie lower and more compressed interquartile ranges) and better minimum $\sigma^2$ values (\ie the most compressed LMO) when compared to Boys.
Since the LMOs generated by the Boys procedure are the result of minimizing the sum over orbital variances (see Eq.~\eqref{eq:boys}), it is again interesting to note that the non-iterative \SCDMG procedure generates LMOs with comparable mean and median $\sigma^2$ values.
In the same breath, we also note that the set of LMOs generated by the \SCDMG procedure includes orbitals that are more diffuse (\ie larger $\sigma^2$ values) when compared to the Boys LMOs; this finding results from the differential treatment of $\sigma$ and $\pi$ orbitals during the \SCDMG and Boys procedures, and will be discussed in more detail below in Sec.~\ref{sec:interpret}.
When considering the bottom panel of Fig.~\ref{fig:ane_ene}, one again sees that the \SCDMG, Boys, and PM schemes produce LMOs that are comparable (on average) with some notable similarities and differences; for instance, the \SCDMG and Boys LMOs have similar bulk locality and maximum $(ii|ii)$ values (both of which are worse than PM), while the \SCDMG and PM LMOs have similar minimum $(ii|ii)$ values that are worse than Boys.

\subsection{Chemical Interpretation of SCDM Orbitals \label{sec:interpret}}

%
%
\begin{figure}[t!]
    \centering
    \includegraphics[width=1.0\linewidth]{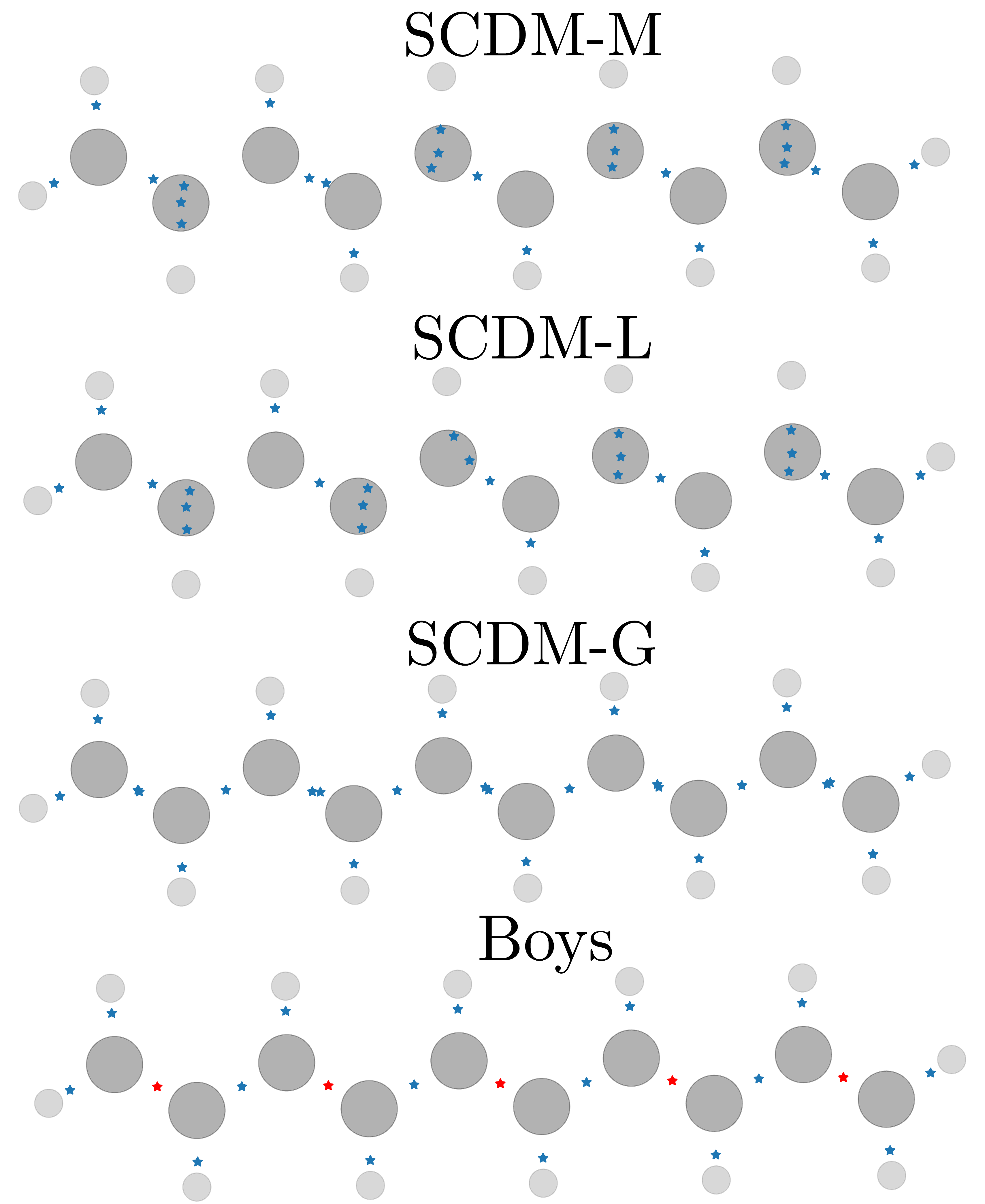}
    \caption{
    Orbital centroid positions ($C_i = \braket{ \phi_i | \hat{\mathbf{r}} | \phi_i}$, blue and red stars) projected onto the molecular plane of the \ce{C10H12} s-\textit{trans} alkene for the LMOs generated using the \SCDMM, \SCDML, \SCDMG, and Boys methods at the HF/cc-pVTZ level of theory.
    For the Boys method, the red stars are used to indicate the location of two equivalent $\tau$ (banana/bent) LMOs.
    }
    \label{fig:centers}
\end{figure}
%
%
In the previous sections, we have demonstrated that LMOs constructed according to the non-iterative SCDM variants introduced in this work are indeed local for molecular systems of varying size, dimensionality, and saturation level.
In many cases (particularly those produced \via \SCDMG), the LMOs are comparable in locality to those produced with iterative schemes which seek to optimize a pre-defined cost function (\eg Boys, PM, etc.).
However, one aspect in which the SCDM variants may differ amongst themselves as well as other localization methods lies in the chemical interpretation of the generated LMOs. 
While most localization methods generate LMOs with qualitative chemical meaning, \eg bonding and lone pairs, the SCDM variants that are restricted to select primarily atom-centered proto-LMOs, \ie \SCDMM and \SCDML, generally do not.
As illustrated in the top two panels of Fig.~\ref{fig:centers}, the LMOs produced by \SCDMM and \SCDML for the \ce{C10H12} s-\textit{trans} alkene are noticeably more atom-centered than their Boys and \SCDMG counterparts.
This is not an entirely unexpected observation as the \SCDMM and \SCDML procedures must select either PAOs or POAOs---most (but not all) of which are atom-centered functions---when forming their respective initial sets of proto-LMOs.
Since the PAOs or POAOs that are not atom-centered tend to have less overlap with the occupied space (and tend to be more delocalized), these functions are less likely to be selected by the \SCDMM or \SCDML procedures, thereby resulting in a primarily atom-centered set of LMOs.
As such, the use of PAOs or POAOs as the underlying proto-LMO basis does not seem to be flexible enough (in general) to generate LMOs that are consistent with standard chemical (\ie Lewis structure) interpretations. 
In contrast, we find that the inherently more flexible \SCDMG procedure generates LMOs for \ce{C10H12} that are bond-centered, with orbital centroid positions distributed similarly to those produced using the iterative Boys procedure (see the bottom two panels of Fig.~\ref{fig:centers}).
In fact, the only qualitative difference between these two LMO sets lies in their description of the double bonds in this alkene system.
We note in passing that the points selected by the pivoted QR factorization during the \SCDMG procedure will depend on the underlying real-space grid; however, we find that the centroids of the final LMOs are qualitatively similar across a wide range of grid densities (see Fig.~\ref{fig:grid2}). 
In other words, the centroids of the \SCDMG LMOs are robust---even in cases where different real-space grids are used and the QR procedure cannot select exactly the same grid points (\eg coarser grids that are not necessarily subsets of finer grids). 

As mentioned above, it is well-known that the Boys and ER localization schemes mix $\sigma$ and $\pi$ bonds in systems containing multiple (\ie double and triple) bonds to produce $\tau$ (banana/bent) bonds,~\cite{Boy66} while other methods, particularly those that are based upon the density matrix (\eg PM),~\cite{Pip89} tend to preserve $\sigma\text{-}\pi$ separation.
In this regard, further analysis of \SCDMG shows that this localization scheme seems to fall within this second group of methods as it generates LMOs which qualitatively preserve $\sigma\text{-}\pi$ separation.
%
%
\begin{figure}[t!]
    \centering
    \includegraphics[width=1.0\linewidth]{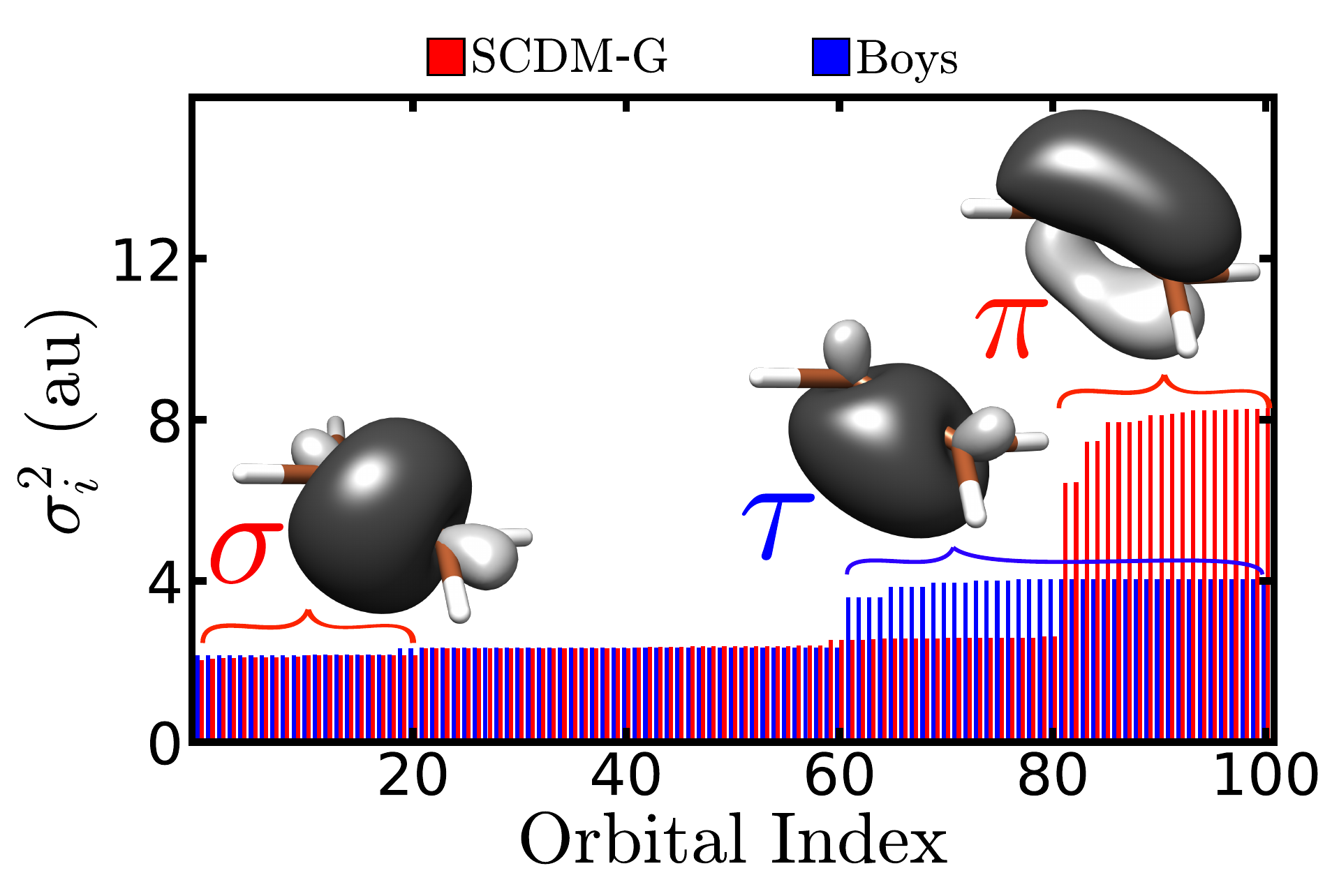}
    \caption{
    Plot of orbital locality (measured \textit{and} sorted by orbital variance) for the LMOs of the \ce{C40H42} s-\textit{trans} alkene generated using the \SCDMG and Boys methods at the HF/cc-pVTZ level of theory.
    In the \SCDMG case, the first and last $20$ LMOs (\ie those with the smallest and largest orbital variances, respectively) correspond to the \ce{C=C} double bonds in \ce{C40H42}, with the first (last) set having significant $\sigma_{\rm C=C}$ ($\pi_{\rm C=C}$) character.
    On the other hand, the last $40$ LMOs generated by the Boys scheme have significant $\tau_{\rm C=C}$ character with orbital variances intermediate to the $\sigma_{\rm C=C}$ and $\pi_{\rm C=C}$ \SCDMG LMOs. 
    All other (unlabelled) orbitals correspond to LMOs with significant $\sigma_{\rm C-C}$ or $\sigma_{\rm C-H}$ character.
    \textit{Insets:} For illustrative purposes, the \SCDMG-generated LMOs with significant $\sigma_{\rm C=C}$ and $\pi_{\rm C=C}$ character (as well as the Boys-generated LMOs with significant $\tau_{\rm C=C}$ character) are also included for the ethylene molecule.
    }
    \label{fig:sigpi}
\end{figure}
%
%
This is depicted in Fig.~\ref{fig:sigpi}, which compares the LMOs generated by the \SCDMG and Boys methods for the \ce{C40H42} s-$trans$ alkene.
In this figure, the LMOs have been sorted (in increasing order) according to their orbital variances (or second central moments), from which one can immediately see that LMOs of similar character (\ie $\sigma$, $\pi$, $\tau$) can be clearly distinguished based on clustered $\sigma_i^2$ values.
For example, the most local LMOs produced by the \SCDMG procedure fall within a group of $20$, which visually correspond to the $\sigma$-like components of the $20$ C=C double bonds ($\sigma_{\rm C=C}$) in \ce{C40H42}, while the $20$ most delocalized LMOs visually correspond to the $\pi_{\rm C=C}$ bonds.
In contrast, the Boys method produces $40$ nearly equivalent LMOs with significant $\tau$-character ($\tau_{\rm C=C}$), \ie LMOs of mixed $\sigma$- and $\pi$-character with orbital variances between that of the \SCDMG-generated $\sigma_{\rm C=C}$ and $\pi_{\rm C=C}$ LMOs.
Here, we note in passing that the \SCDMG procedure---which selects a \textit{single} real-space grid point (instead of groups of symmetrically-equivalent grid points)---does not explicitly account for the underlying symmetry of the molecular system.
As such, $\sigma\text{-}\pi$ separation is only qualitatively preserved by the \SCDMG procedure, and this is clearly illustrated by the insets in Fig.~\ref{fig:sigpi} which demonstrate that the \SCDMG-generated LMOs for the planar ($D_{\rm 2h}$) ethylene molecule have significant (but not perfect) $\sigma_{\rm C=C}$ and $\pi_{\rm C=C}$ character.

In fact, this classification of \SCDMG-generated LMOs into $\sigma_{\rm C=C}$ and $\pi_{\rm C=C}$ orbital sets is also supported by the clustering observed in the orbital expectation values of the Fock operator in Table~\ref{tab:energy}.
Defined as $F_{ii} \equiv \braket{\phi_i|\hat F|\phi_i}$, these quantities are analogous to the orbital energy eigenvalues in the CMO basis (\ie $\epsilon_{i} = \braket{\psi_i|\hat F|\psi_i}$), and can be used as a semi-quantitative measure of the ``energy" of a given LMO.
Considering the minimum, mean, and maximum $F_{ii}$ values provided in Table~\ref{tab:energy}, one can immediately see that each orbital set (\ie \SCDMG $\sigma_{\rm C=C}$, \SCDMG $\pi_{\rm C=C}$, Boys $\tau_{\rm C=C}$) is characterized by a narrow but distinct range of expectations values.
In addition, one can also observe a clear ``energetic'' separation between the \SCDMG $\sigma_{\rm C=C}$ and $\pi_{\rm C=C}$ orbital sets, while the Boys $\tau_{\rm C=C}$ LMOs are characterized by intermediate (and very homogeneous) $F_{ii}$ values.
Here, we would argue that this ``energetic'' separation between \SCDMG-generated LMOs (in conjunction with their local $\sigma$- and $\pi$-like character) might prove beneficial when choosing an active space and/or truncating the number of excitations during post-HF electron correlation methods.
%
%
\begin{table}[t!]
    \centering
    \caption{
    Expectation values of the Fock operator, $F_{ii} = \braket{\phi_i|\hat F|\phi_i}$, corresponding to each LMO set described in Fig.~\ref{fig:sigpi}.
    Minimum, mean, and maximum values for $F_{ii}$ (in Hartrees) indicate a clear energetic separation between the \SCDMG-generated LMOs with significant $\sigma_{\rm C=C}$ and $\pi_{\rm C=C}$ character, as well as the intermediate values of the Boys-generated LMOs with significant $\tau_{\rm C=C}$ character. 
    }
    \begin{tabular}{c | c c c c}
    \hline\hline
    Method & Orbital Set        &  Minimum &   Mean & Maximum \\
    \hline
    \SCDMG & $\sigma_{\rm C=C}$ &  $-0.786$ & $-0.775$ & $-0.756$  \\
    \SCDMG & $\pi_{\rm C=C}$    &  $-0.492$ & $-0.478$ & $-0.464$  \\
    Boys   & $\tau_{\rm C=C}$   &  $-0.625$ & $-0.624$ & $-0.618$  \\
    \hline\hline
    \end{tabular}
    \label{tab:energy}
\end{table}
%
%

This contrasting treatment of the C=C double bonds in extended alkenes by the \SCDMG and Boys schemes can also be used to explain the differences observed in the orbital locality metrics in Fig.~\ref{fig:ane_ene}.
Since the \SCDMG $\pi_{\rm C=C}$ LMOs are more diffuse than the Boys $\tau_{\rm C=C}$ LMOs (see Fig.~\ref{fig:sigpi}), the increased maximum $\sigma_i^2$ values (and decreased minimum $(ii|ii)$ values) observed for the \SCDMG LMOs can be rationalized as a manifestation of approximate $\sigma\text{-}\pi$ separation (and not an artifact of poor performance) by the non-iterative \SCDMG procedure.
In fact, this also explains the close resemblance between the \SCDMG and PM orbital locality metrics in Fig.~\ref{fig:ane_ene}, as the PM scheme also preserves $\sigma\text{-}\pi$ symmetry~\cite{Pip89}.
%

\subsection{SCDM Orbitals as Initial Guesses for Iterative Localization Methods \label{sec:guess}}

Based on the analysis presented herein, we expect that direct use of the non-iterative \SCDMG (or even \SCDMM/\SCDML) LMOs may suffice in most chemical applications.
Since SCDM-generated LMOs are derived directly from the density matrix and are completely agnostic with respect to any single measure of orbital locality, these LMOs can also be used as an efficient and unbiased starting point (or initial guess) for generating LMOs that have been optimized over a pre-defined metric (\eg Boys, PM, ER).
In this regard, a high-quality initial guess has the potential to reduce the number of iterations in such non-linear optimization schemes, as well as avoid the convergence issues seen in conventional algorithms utilizing pairwise Jacobi rotations (\ie without having to resort to more sophisticated gradient- or Hessian-based optimization protocols).~\cite{hoy12b,Leh13}
%
%
\begin{figure}[t!]
    \centering
    \includegraphics[width=1.0\linewidth]{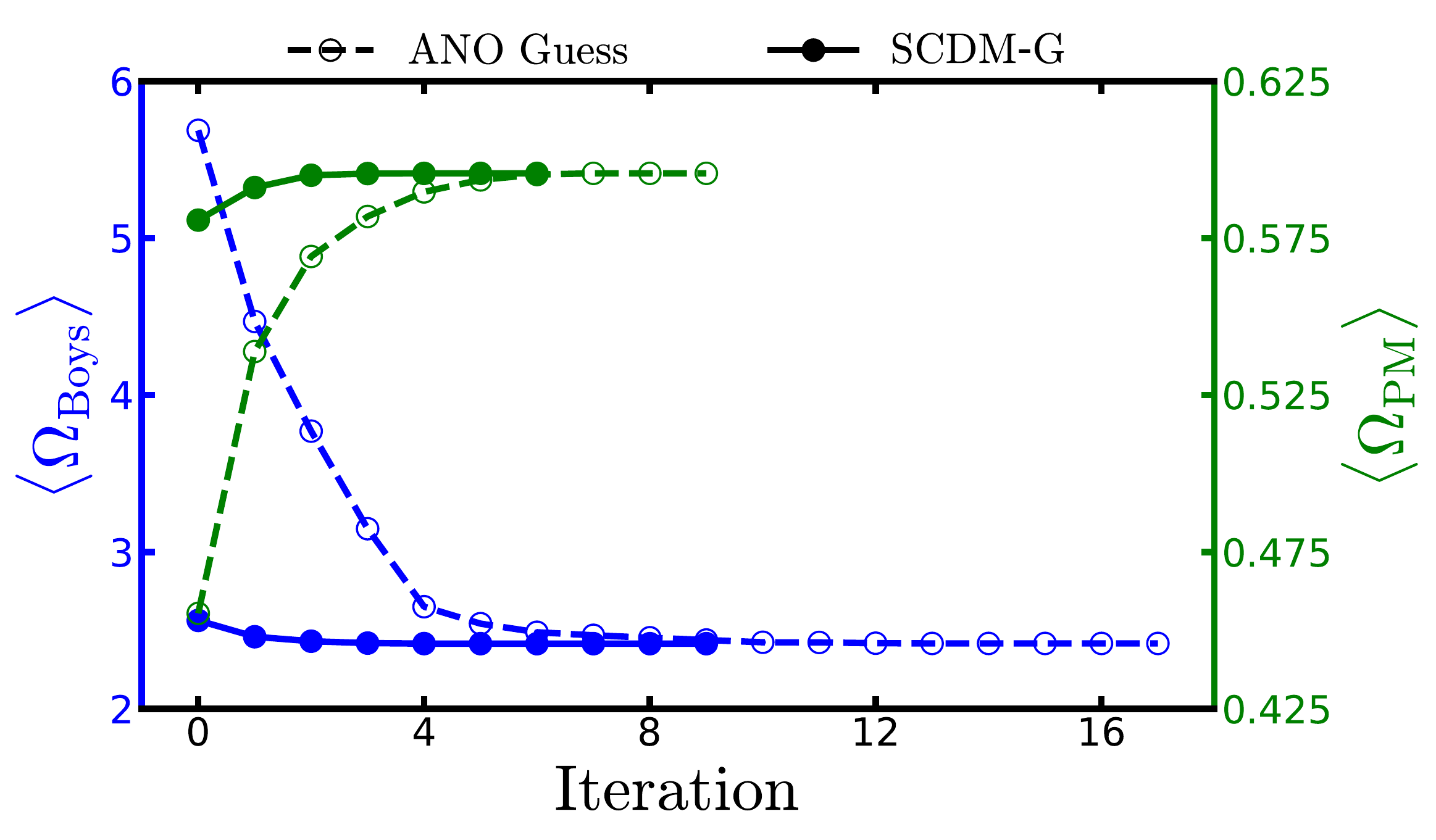} 
    \caption{
    Convergence profiles of the iterative Boys (left, blue) and PM (right, green; using Mulliken populations) schemes during LMO generation for enalapril at the HF/cc-pVTZ level of theory.
    Closed circles indicate iterative procedures that have been initialized with \SCDMG-generated LMOs, while open circles indicate initialization with the (default) atomic natural orbital (ANO) based guess in \texttt{PySCF}.
    For ease of comparison to the other figures in this work, the Boys and PM metrics in Eqs.~\eqref{eq:boys}--\eqref{eq:pm} have been averaged over the LMOs, \ie $\braket{\Omega_{\rm Boys}} = \Omega_{\rm Boys}/N_{\rm occ}$.
    }
    \label{fig:guess}
\end{figure}
%
%
Fig.~\ref{fig:guess} indeed supports these claims by showing that the non-iterative \SCDMG orbitals are simultaneously characterized by nearly converged $\Omega_{\rm Boys}$ and $\Omega_{\rm PM}$ cost functions, and are therefore excellent starting points for these iterative orbital localization schemes.
In fact, starting either of these optimization schemes with \SCDMG-generated LMOs as the initial guess reduced the required number of iterations by $30\%\mathrm{-}50\%$ for the drug molecule enalapril (when compared to the default projected atomic natural orbital (ANO)~\cite{wid90,roo04} guess in \texttt{PySCF}).
For larger and/or more complicated/pathological systems where a good initial guess is crucial for efficient convergence to the minimum/maximum, we expect that this effect will be even more pronounced.
Even for small (and relatively simple) molecular systems such as water, we have seen the iterative Boys orbital localization scheme (with the default ANO initial guess in \texttt{PySCF}) converge to a local minimum; in this case, the use of \SCDMG orbitals as the initial guess resulted in rapid convergence to a set of LMOs with an $\Omega_{\rm Boys}$ cost function that was nearly $20\%$ smaller.

\section{Conclusions and Future Outlook  \label{sec:conclusion}}

In this work, we have extended the recently proposed selected columns of the density matrix (SCDM) methodology---a non-iterative and real-space procedure for generating localized occupied orbitals for condensed-phase systems---to the construction of local molecular orbitals (LMOs) in systems described using non-orthogonal atomic orbital (AO) basis sets.
In doing so, we have presented three different theoretical and algorithmic variants of SCDM (\SCDMM, \SCDML, and \SCDMG) that can be used in conjunction with an underlying AO basis set, \ie the standard approach for performing high-level quantum chemical calculations on molecular systems.
In doing so, we have explicitly shown that the \SCDMM and \SCDML variants (which are based on a pivoted QR factorization of the well-known Mulliken and L{\"o}wdin representations of the density matrix) are tantamount to selecting a well-conditioned set of projected atomic orbitals (PAOs) and projected (symmetrically-) orthogonalized atomic orbitals (POAOs), respectively, as proto-LMOs that can be orthogonalized to exactly span the occupied space.
As such, the \SCDMM and \SCDML procedures might find use in choosing a well-conditioned (or orthogonal) set of PAO- or POAO-based LMOs that exactly span the occupied space and could potentially be used in local WFT and DFT methods.
The \SCDMG variant, which is based on a real-space grid representation of the wavefunction (\via a transformation from the AO representation), has the added flexibility of considering a larger number of grid points (or $\delta$ functions) when selecting a set of well-conditioned proto-LMOs.

With these SCDM variants in hand, we performed a detailed comparative analysis of their performance for molecular systems of varying size, dimensionality, and saturation level.
In doing so, we found that all three SCDM variants produced LMOs that are significantly more local than the redundant and non-orthogonal PAO (or POAO) representation of the occupied space.
In many cases, the LMOs generated by the direct/non-iterative SCDM approach (in particular those produced with the more flexible \SCDMG variant) are quite comparable in orbital locality to those produced with the iterative Boys or PM localization schemes, while still remaining completely agnostic towards any particular (user-defined) orbital locality metric and/or initial guess. 
Although all three SCDM variants are based on the density matrix, only the grid-based \SCDMG procedure (like PM) generates LMOs that qualitatively preserve $\sigma\text{-}\pi$ separation (in systems like the s-\textit{trans} alkenes) and can be interpreted using standard chemical concepts.
This preservation of $\sigma\text{-}\pi$ symmetry may also prove to be computationally relevant, as local electron correlation methods based on \SCDMG LMOs can exploit the physical and energetic separation of the $\sigma$ and $\pi$ LMOs, which is an option that is not available with the mixed $\tau$ (banana/bent) LMOs produced by the Boys or ER schemes.

Our findings strongly suggest that the LMOs generated by the three non-iterative/direct SCDM variants introduced herein are robust, comparable in orbital locality to those produced with the iterative Boys or PM localization schemes, and completely agnostic towards any single locality metric.
While SCDM-generated LMOs should suffice in most chemical applications, we have also briefly explored the use of these orbitals as an unbiased and cost-effective initial guess for such popular localization schemes.
Although the convergence analysis presented in this work is necessarily limited and far from complete, our findings suggest that the use of \SCDMG-generated LMOs as an initial guess has the potential to improve the convergence of iterative orbital localization schemes, and we expect such issues to be even more pronounced when treating large and complex molecular systems.
As such, we recommend that the SCDM variants introduced herein should be implemented into the quantum chemistry community codes, as both a standalone option for novel LMO generation and an initial guess option for iterative orbital localization.
Although the \SCDMG procedure requires a real-space grid (in addition to the underlying AO basis set), we stress here that a dense grid (particularly in regions of small electron density) is not required to generate high-quality and robust LMOs.
In this regard, the increased computational scaling associated with the grid-based \SCDMG variant (\eg $\mathcal{O}(N_{\rm occ}^2 N_{\rm grid})$ versus $\mathcal{O}(N_{\rm AO}^3)$ for \SCDMM and \SCDML) should be largely ameliorated using stochastic sampling techniques.~\cite{Dam17b} 
Such algorithmic improvements will become particularly important when dealing with large and complex molecular systems containing $100\text{s}\mathrm{-}1000\text{s}$ of atoms, and should be considered when implementing this method into a given codebase.\\



\section*{Acknowledgements}

All authors acknowledge partial financial support from Cornell University through start-up funding.
This material is based upon work supported by the National Science Foundation under Grant No.\ CHE-1945676.
RAD also gratefully acknowledges financial support from an Alfred P. Sloan Research Fellowship.
This research used resources of the National Energy Research Scientific Computing Center, which is supported by the Office of Science of the U.S. Department of Energy under Contract No.\ DE-AC02-05CH11231.

\bibliographystyle{achemso-title.bst}
\bibliography{references}

\begin{mcitethebibliography}{117}
\bibitem[Not()]{NoteCost}
One could also compute the QR factorization step in the \SCDMM procedure in an
  analogous manner to Eq.~\eqref{eq:qr_psi}, which would scale as
  $\mathcal{O}(N_{\rm AO} N_{\rm occ}^2)$ and therefore be preferred for larger
  systems and/or basis sets. Here, we outline the more transparent approach
  that clearly illustrates the connections between the selected columns of
  $\mathbf{P}\mathbf{S}$ (\ie the proto-LMOs) and the PAOs.\relax
\mciteBstWouldAddEndPunctfalse
\mciteSetBstMidEndSepPunct{\mcitedefaultmidpunct}
{}{\mcitedefaultseppunct}\relax
\EndOfBibitem
\end{mcitethebibliography}

\newpage
\appendix

\section{Grid Dependence of \SCDMG}
\label{sec:grid_appendix}

\begin{figure}[th!]
    \centering
    \includegraphics[width=.99\linewidth]{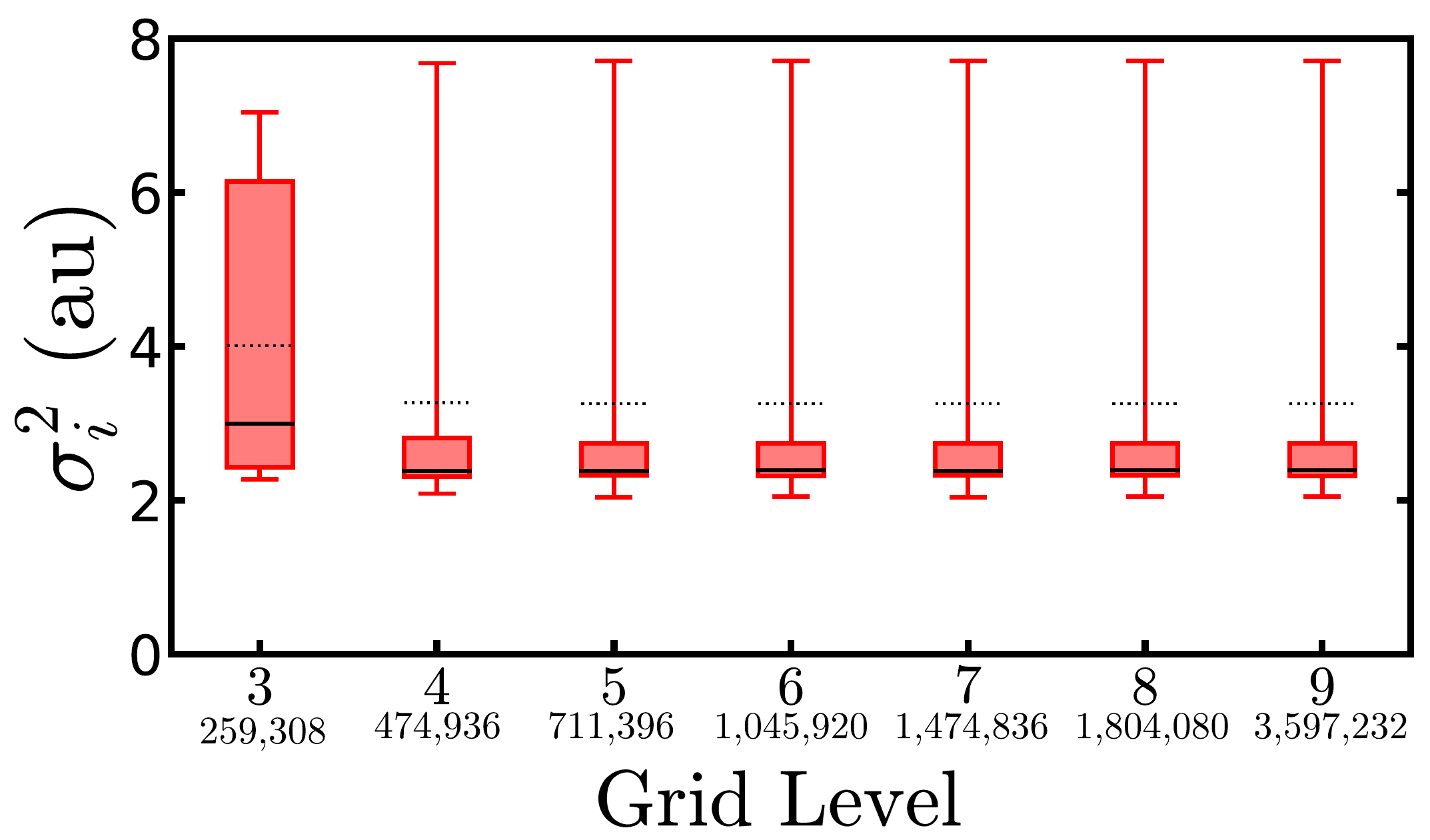}
    \caption{
    Box-and-whisker plots of orbital locality (as measured by the orbital variance or second central moment) versus grid level for the \ce{C10H12} s-\textit{trans} alkene. 
    All LMOs were generated with the \SCDMG procedure at the HF/cc-pVTZ level of theory, and the grid levels correspond to those used in \texttt{PySCF} (with the total number of grid points after pruning provided below each level).
    In each case, the whiskers extend to the minimum and maximum $\sigma_{i}^{2}$ values, while the box demarcates the $Q_1\mathrm{-}Q_3$ ($25\%\mathrm{-}75\%$) interquartile range; the solid (dashed) lines within each box mark the position of the median (mean).
    }
    \label{fig:grid1}
\end{figure}

In order to investigate how the underlying real space grid affects the LMOs produced by the \SCDMG procedure, we have performed \SCDMG with Treutler--Ahlrichs--Lebedev~\cite{tre95,leb1999} grids of varying sizes for the \ce{C10H12} s-\textit{trans} alkene. 
As plotted in Fig.~\ref{fig:grid1}, the locality of the LMOs produced by the \SCDMG procedure is rather insensitive to the size of the underlying real space grid, yielding nearly identical  $\sigma_i^2$ distributions for grid levels $\ge 4$ (a medium-quality grid containing $(50,302)$ [H] and $(75,302)$ [C] radial and angular grid points before pruning) in the \texttt{PySCF} electronic structure package.~\cite{sun18} 
We note in passing that grid level $9$ was used throughout this work ($(200,1454)$ [H,C] before pruning, the largest grid level in \texttt{PySCF}) to ensure that grid effects were negligible; however, Fig.~\ref{fig:grid1} shows that \SCDMG orbitals of nearly identical quality can be obtained using grid levels $\ge 4$, which are nearly an order of magnitude smaller.
We also note in passing that one can further increase the computational efficiency of \SCDMG by randomly sub-sampling the grid prior to performing the column-pivoted QR factorization.~\cite{Dam17b}

\begin{figure*}[ht!]
    \centering
    \includegraphics[width=0.8\linewidth]{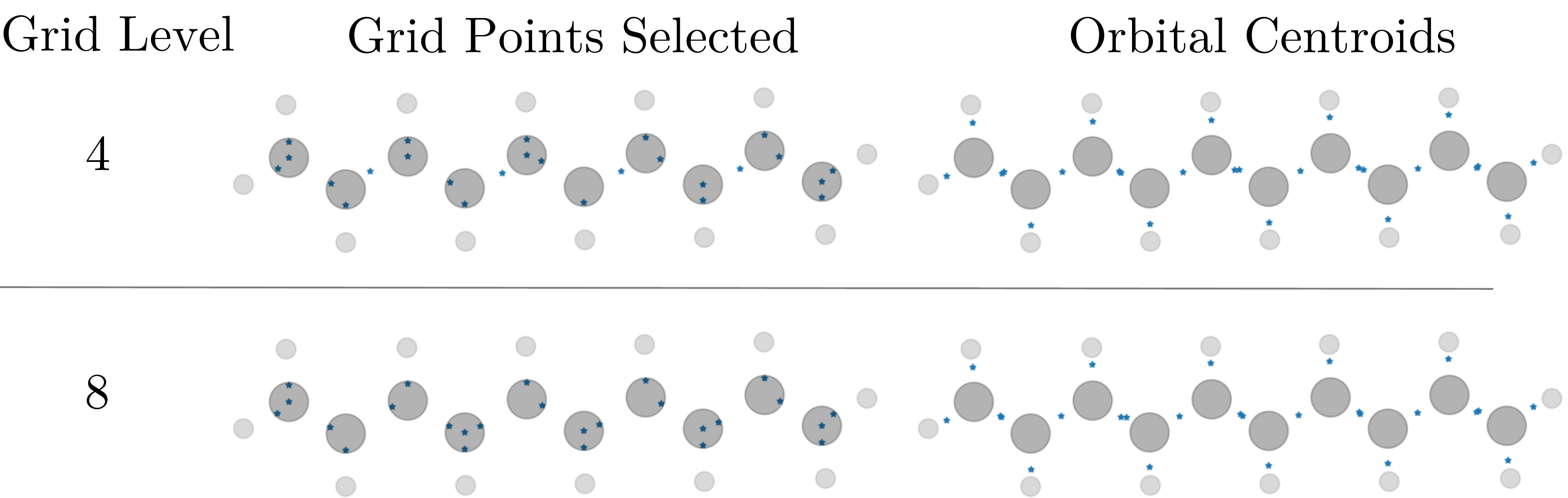}
    \caption{
    Positions of the grid points selected by the QR procedure and resulting \SCDMG LMO centroids projected onto the molecular plane of the \ce{C10H12} s-\textit{trans} alkene.
    All LMOs were generated with the \SCDMG procedure at the HF/cc-pVTZ level of theory, and the grid levels correspond to those used in \texttt{PySCF} (see Fig.~\ref{fig:grid1}).
    }
    \label{fig:grid2}
\end{figure*}

As depicted in Fig.~\ref{fig:grid2}, the points selected by the QR factorization will depend on the underlying real space grid; however, the centroids of the final \SCDMG LMOs tend to be qualitatively similar between grid levels $4$ and $8$, and visually indistinguishable between grid levels $8$ and $9$ (\cf Fig.~\ref{fig:centers} in the main text).
This finding is particularly encouraging as many grid generation schemes (such as the one used here, which is standard in molecular quantum mechanics codes) do not define larger grid levels as strict supersets of smaller grid levels.
In other words, the centroids of the \SCDMG LMOs are robust---despite the fact that the QR procedure is not able to select the same \textit{exact} grid points when used with different grid levels.

\section{Basis Set Dependence of \SCDMM and \SCDML}
\label{sec:basis_appendix}

In Sec.~\ref{sec:prop}, we showed that the \textit{locality} of the \SCDMM and  \SCDML LMOs has a weak dependence on the choice of basis within the cc-pVXZ and aug-cc-pVXZ (with X = D, T, Q) basis set family.
As depicted in Fig.~\ref{fig:basis}, we demonstrated that the variation in the locality of the \SCDMM and \SCDML LMOs is similar to that found for the CMOs.
However, this observation is in stark contrast to the PAOs and POAOs, whose locality depends strongly on the choice of the underlying basis set.
Since the \SCDMM and \SCDML procedures directly rely on the PAOs and POAOs, respectively, through the choice of proto-LMOs (see Eqs.~(16) and (18) in the main text), the final \SCDMM and \SCDML LMOs implicitly inherit a dependence on the basis set through these projected orbitals.
As such, the \textit{character} of the final \SCDMM and \SCDML LMOs are not necessarily invariant with respect to changes in the basis set.
For example, different proto-LMOs (and hence LMOs) may be produced when the basis functions are rotated with respect to the global coordinate system (see Sec.~\ref{sec:orientation_appendix}) or when different basis set families, \eg those with different contraction schemes, are used.
Before moving on to look at this important theoretical (and potentially practical) issue in more detail, we note that there are settings where a well-conditioned and orthogonal subset of PAOs or POAOs that exactly span the occupied space are desired (\ie for use in local WFT and DFT methods); in such cases, the \SCDMM and \SCDML procedures can be used to directly accomplish this task.

To assess the basis set dependence of the \SCDMM and \SCDML methods, we now consider the valence LMOs produced by these methods for ethylene (\ce{C2H2}) when using three different basis set families.
Here, we do not focus on the typical notion of ``basis set dependence,'' \ie the convergence of a given method with respect to changes in the basis set due to the addition of higher-order polarization and/or diffuse functions to approach the complete basis set limit.
Instead, we focus our discussion on how the final \SCDMM and \SCDML LMOs depend on changes to the contraction schemes used to construct basis sets of comparable (\ie triple-$\zeta$) quality, as these more subtle changes will lead to distinct sets of PAOs and POAOs without drastically changing the occupied space.
Specifically, we investigate the locality and character of the \SCDMM, \SCDML, and \SCDMG LMOs when employing the cc-pVTZ~\cite{dun89} and ANO-RCC-VTZP\cite{wid90,roo04} basis sets (which are constructed using general contractions) and the def2-TZVPP\cite{wei05} basis set (which are constructed using segmented contractions).

\begin{figure*}[ht!]
    \centering
    \includegraphics[width=.32\linewidth]{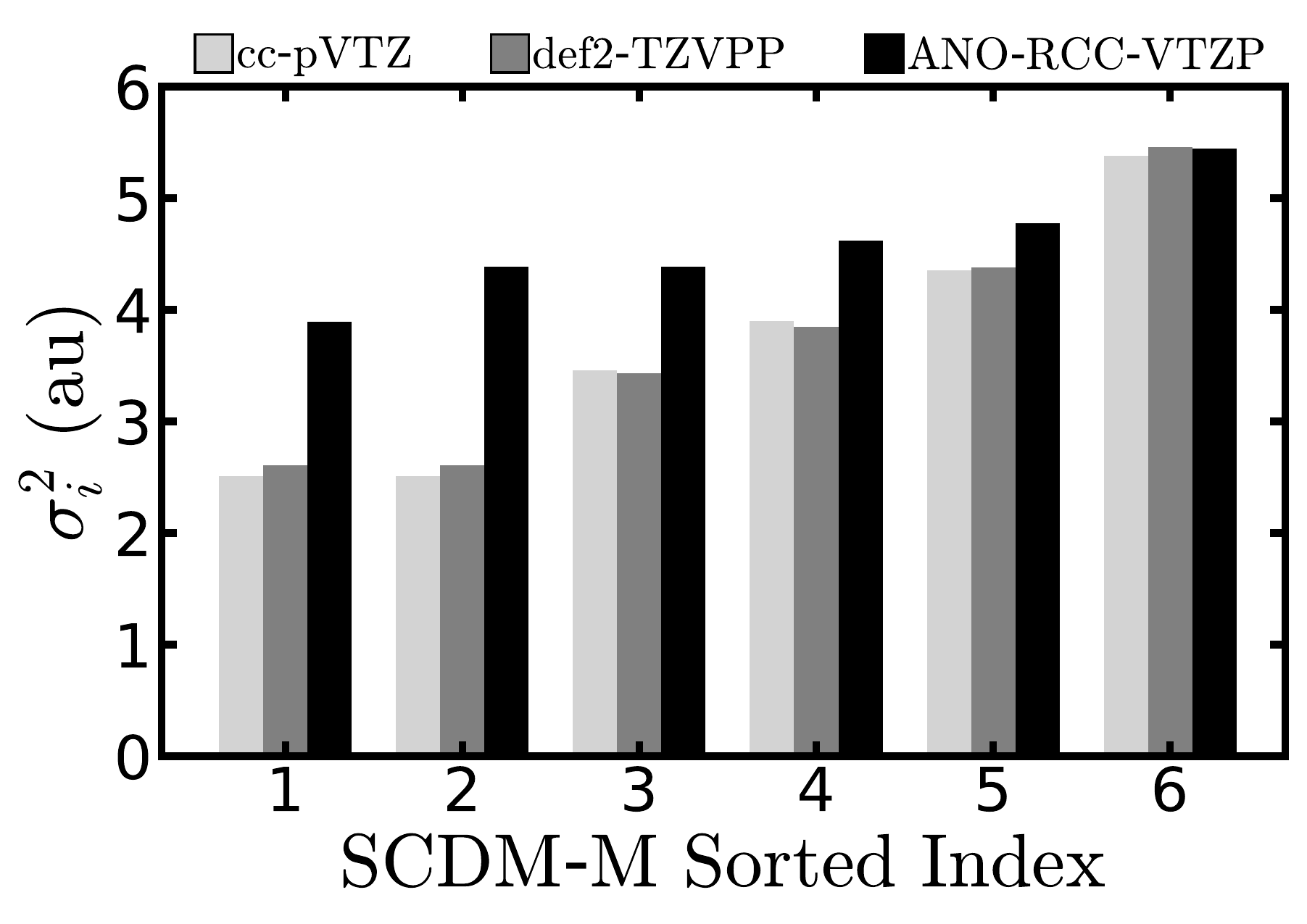}
    \includegraphics[width=.32\linewidth]{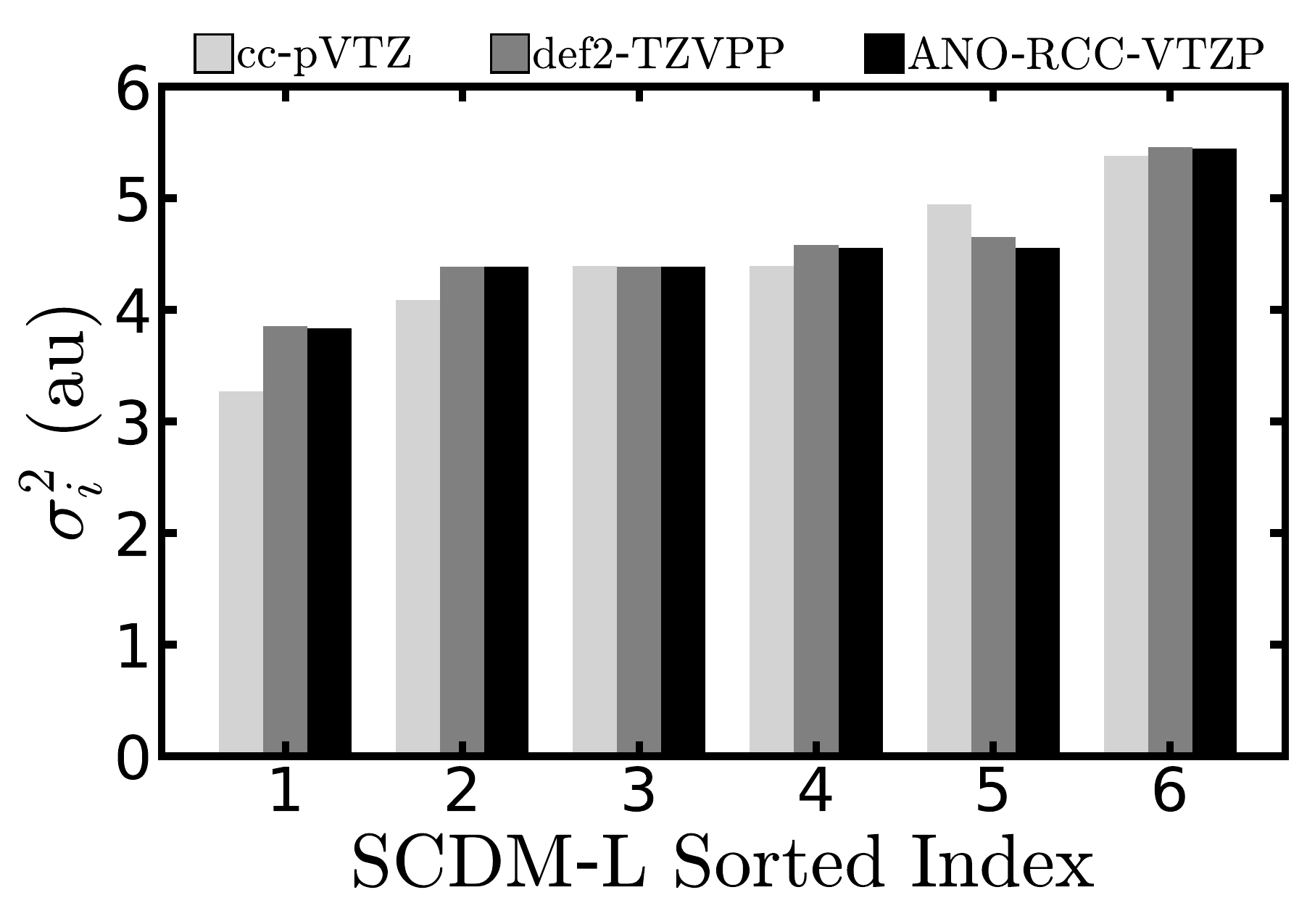}
    \includegraphics[width=.32\linewidth]{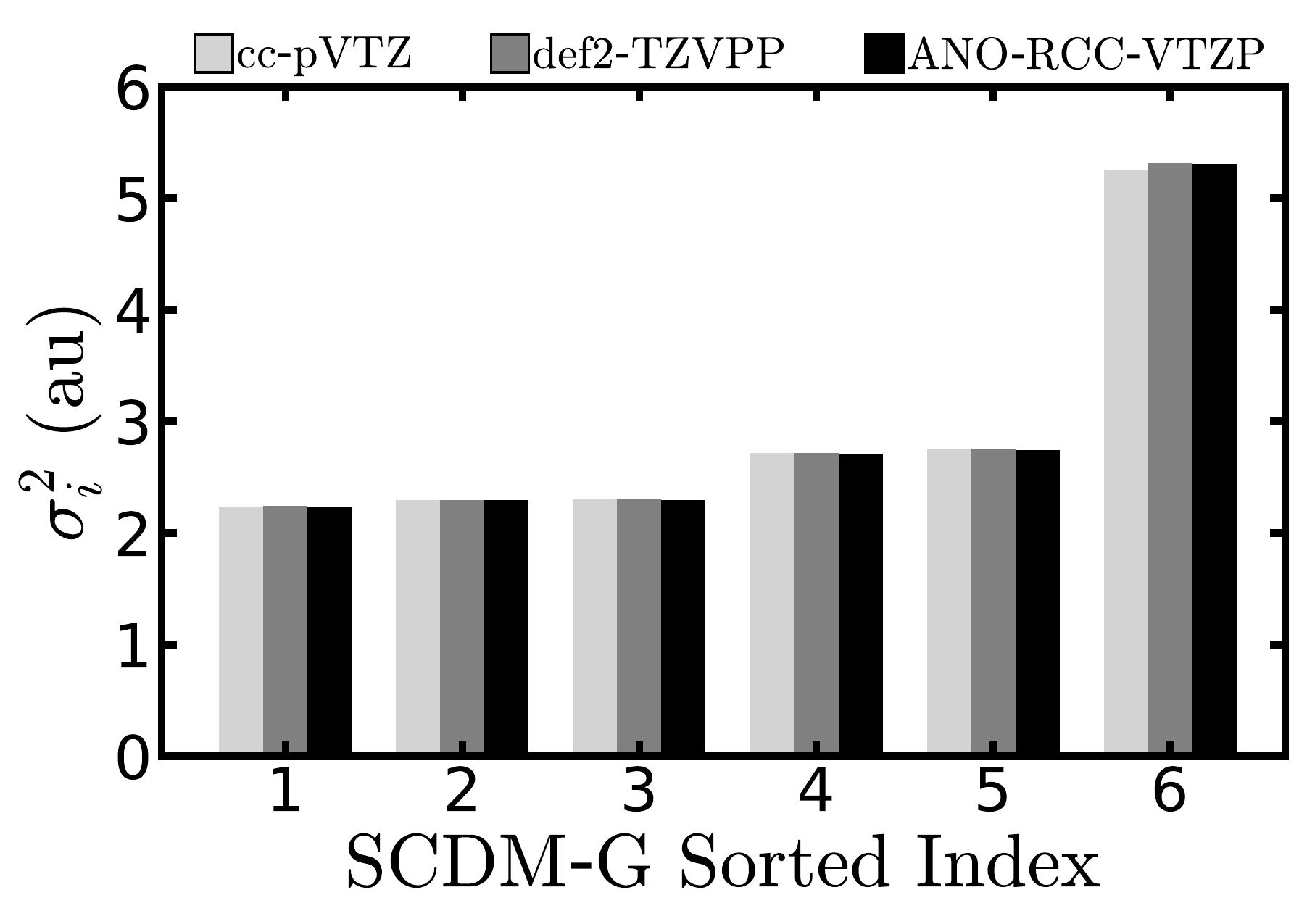}
    \caption{
    Plot of orbital locality (measured and sorted by orbital variance) for the valence LMOs of ethylene (\ce{C2H4}) generated using the \SCDMM (\textit{left}), \SCDML (\textit{middle}), and \SCDMG (\textit{right}) methods at the HF/cc-pVTZ, HF/def2-TZVPP, and HF/ANO-RCC-VTZP levels of theory.
    Since \SCDMM and \SCDML directly rely on the PAOs and POAOs through the choice of proto-LMOs, the locality of the LMOs produced by these schemes is not invariant to changes in the basis set contraction scheme.
    In contrast, the locality of the LMOs produced by the grid-based \SCDMG method is essentially invariant to such changes in the basis set.
    }
    \label{fig:basis_appendix}
\end{figure*}
\begin{figure*}[ht!]
    \centering
    \includegraphics[width=0.7\linewidth]{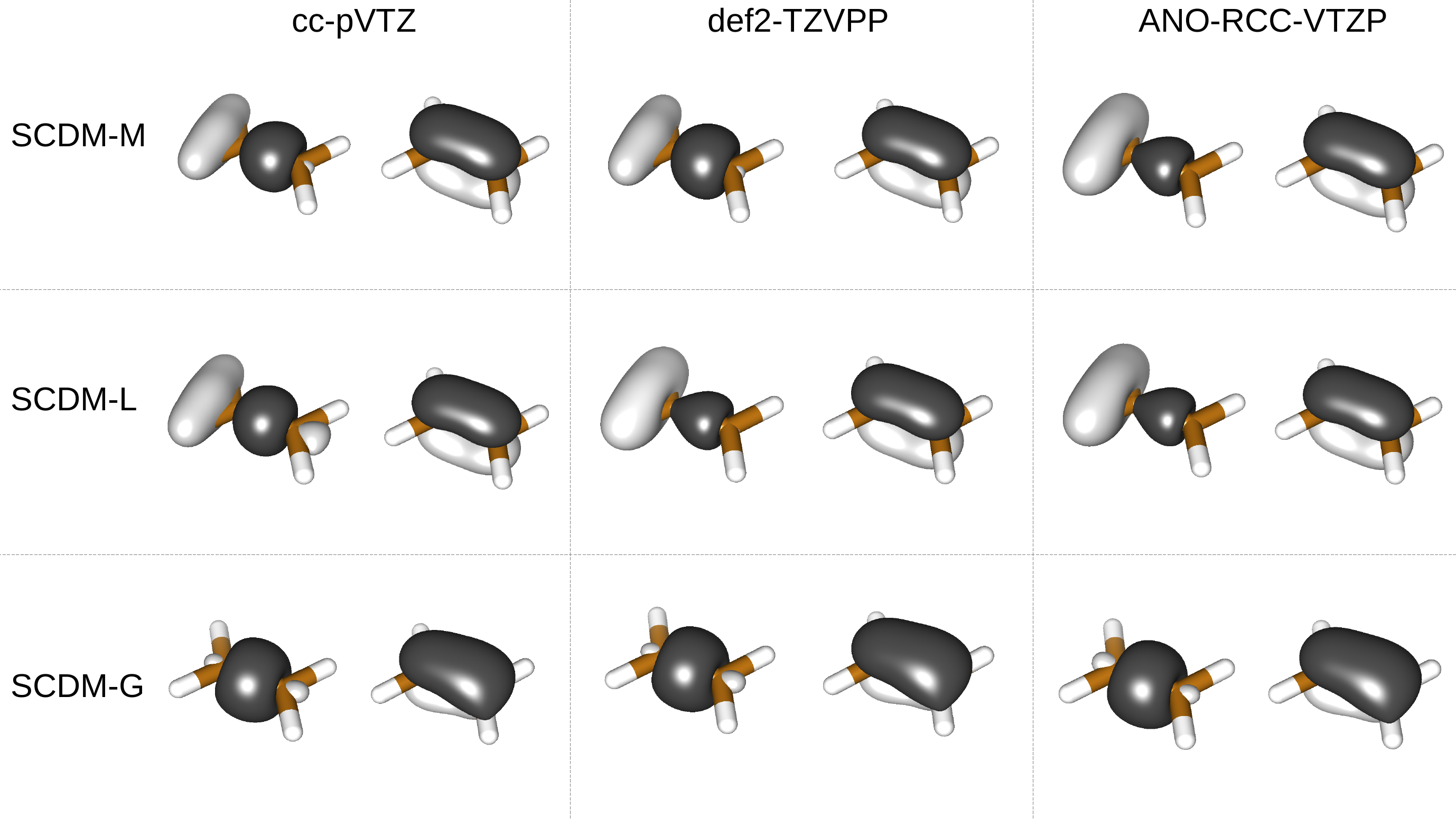}
    \caption{
    Graphical depiction of the two LMOs centered around the \ce{C=C} bond in ethylene (\ce{C2H4}) generated using the \SCDMM, \SCDML, and \SCDMG methods at the HF/cc-pVTZ, HF/def2-TZVPP, and HF/ANO-RCC-VTZP levels of theory.
    Here, we find that the LMOs produced by \SCDMM and \SCDML show some variability with the basis set contraction scheme, although they maintain their characteristic shape and morphology. 
    In contrast, the \SCDMG LMOs are essentially independent of the chosen basis set.
    }
    \label{fig:basis_orbitals}
\end{figure*}

As depicted in Fig.~\ref{fig:basis_appendix}, the locality of the final \SCDMM and \SCDML LMOs can exhibit some variability with respect to changes in the underlying basis set contraction scheme.
As mentioned above (see Secs.~\ref{sec:extension} and~\ref{sec:prop} in the main text for more details) this dependence originates from the fact that \SCDMM and \SCDML use PAOs and POAOs as proto-LMOs during the localization procedure, and these projected orbitals strongly depend on the contraction scheme used to generate the basis set.
Interestingly, the L\"{o}wdin-based \SCDML method seems to largely temper these differences---a fact that we attribute to the symmetric orthogonalization procedure used to generate the POAO-based proto-LMOs, which mixes the basis functions and thereby effectively blurs the distinction between different contraction schemes.
In Fig.~\ref{fig:basis_orbitals}, we also consider how the character of these LMOs depends on the basis set contraction scheme.
From these plots, one sees only minor variations in the LMOs produced by the \SCDMM and \SCDML methods as the basis set is changed from cc-pVTZ to def2-TZVPP to ANO-RCC-VTZP; in other words, the characteristic shape and morphology of these LMOs is largely invariant to the underlying basis set contraction scheme.

Here we note that localization methods like Boys and \SCDMG---which are fundamentally based on the occupied space and invariant to any non-singular transform of the basis functions---avoid these issues once the occupied CMOs are sufficiently converged.
Since this tends to occur when using triple-$\zeta$ quality basis sets (such as those considered herein), we do not expect to see any significant changes in the character and/or locality of these LMOs.
For \SCDMG, this hypothesis is clearly supported by Figs.~\ref{fig:basis_appendix}--\ref{fig:basis_orbitals}, where we see virtually no change in the \SCDMG LMOs as a function of the basis set used (although the locality and character of the LMOs produced by the more flexible \SCDMG method can differ significantly from those produced by \SCDMM and \SCDML).
As discussed above in Sec.~\ref{sec:grid_appendix}, \SCDMG does have an additional dependence on the real space grid, but the grids employed in this work are completely independent of the basis set used for the underlying mean-field HF calculation.

\section{Molecular Orientation Dependence of \SCDMM and \SCDML}
\label{sec:orientation_appendix}

Since the basis sets most commonly used in molecular quantum mechanics (\ie atom-centered Gaussians) are oriented with respect to a \textit{global} coordinate system, any calculation performed using these basis functions will have an inherent dependence on the orientation (not rigid translations) of the molecule with respect to this coordinate system. 
To avoid this dependence on the coordinate system, most software packages automatically reorient/realign the molecule into its so-called canonical/standard nuclear orientation~\cite{gil93} at the very beginning of the calculation.
If one chooses not to use this convention, the character (and to a lesser extent, the locality) of the LMOs produced by the \SCDMM and \SCDML procedures will also depend on the molecular orientation.
This can again be traced to the dependence of \SCDMM and \SCDML on the PAOs and POAOs, respectively, through the choice of proto-LMOs.
Here, we note in passing that this dependence on the molecular orientation could also be largely eliminated by augmenting the PAOs (POAOs) with \textit{rotated basis functions} that have been projected onto the occupied space, thereby enabling \SCDMM (\SCDML) to choose from a richer and more diverse set of proto-LMOs.

\begin{figure*}[ht!]
    \centering
    {\includegraphics[width=0.32\linewidth]{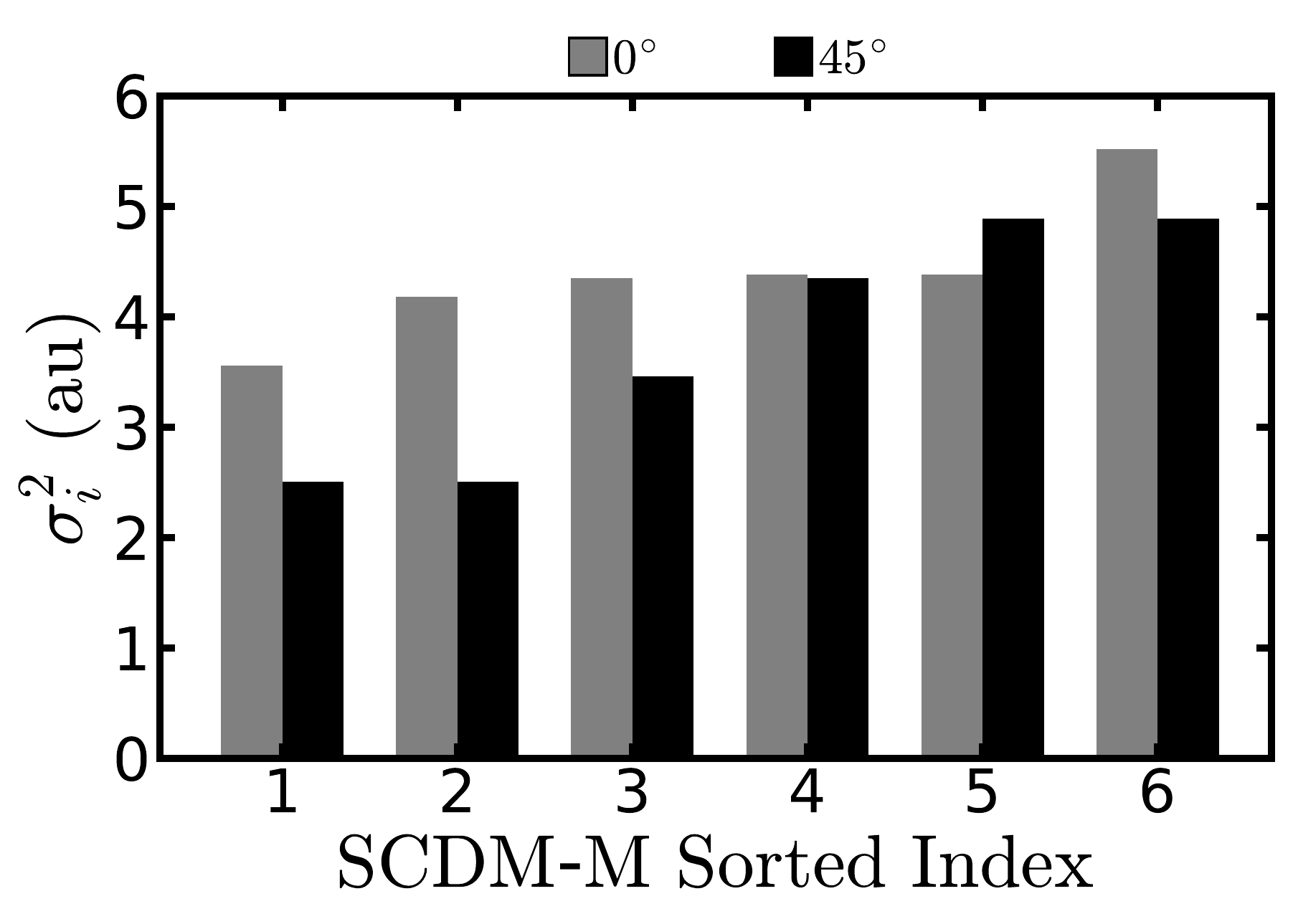} } \hspace{5mm}
    {\includegraphics[width=0.32\linewidth]{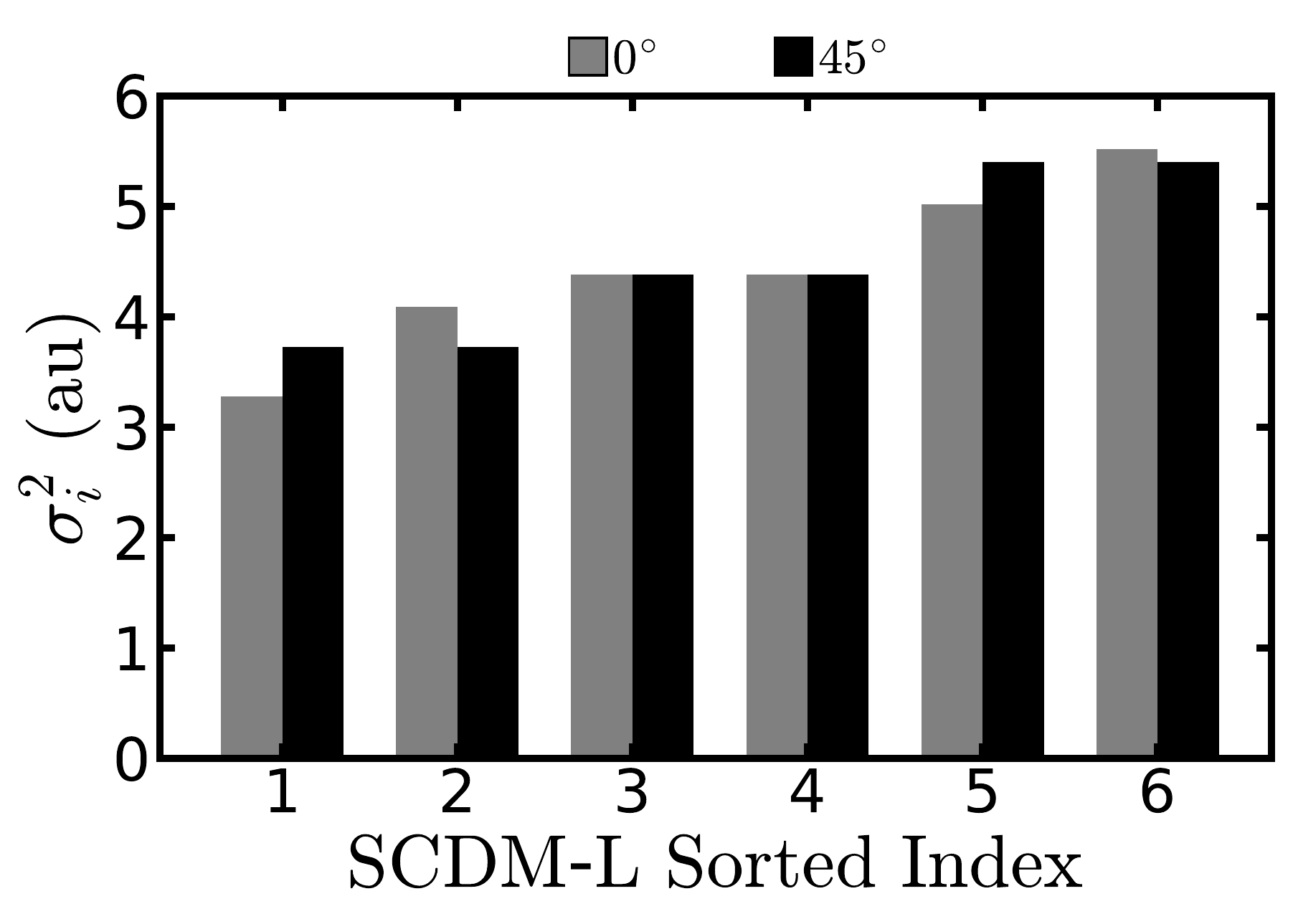} } \\
    \caption{
    Orbital locality (measured and sorted by orbital variance) for the valence LMOs of ethylene (\ce{C2H4}) in two different molecular orientations generated using the \SCDMM (\textit{left}) and \SCDML (\textit{right}) methods at the HF/cc-pVTZ level of theory.
    Here, $0^\circ$ and $45^\circ$ refer to the angle between the molecular plane and the global $yz$ plane (see Fig.~\ref{fig:angle2}).
    As described in the text, failure to use the canonical/standard nuclear orientation may lead to \SCDMM and \SCDML LMOs whose locality depends on the molecular orientation.
    }
    \label{fig:angle}
\end{figure*}

\begin{figure}[ht!]
    \centering
    \includegraphics[width=0.65\linewidth]{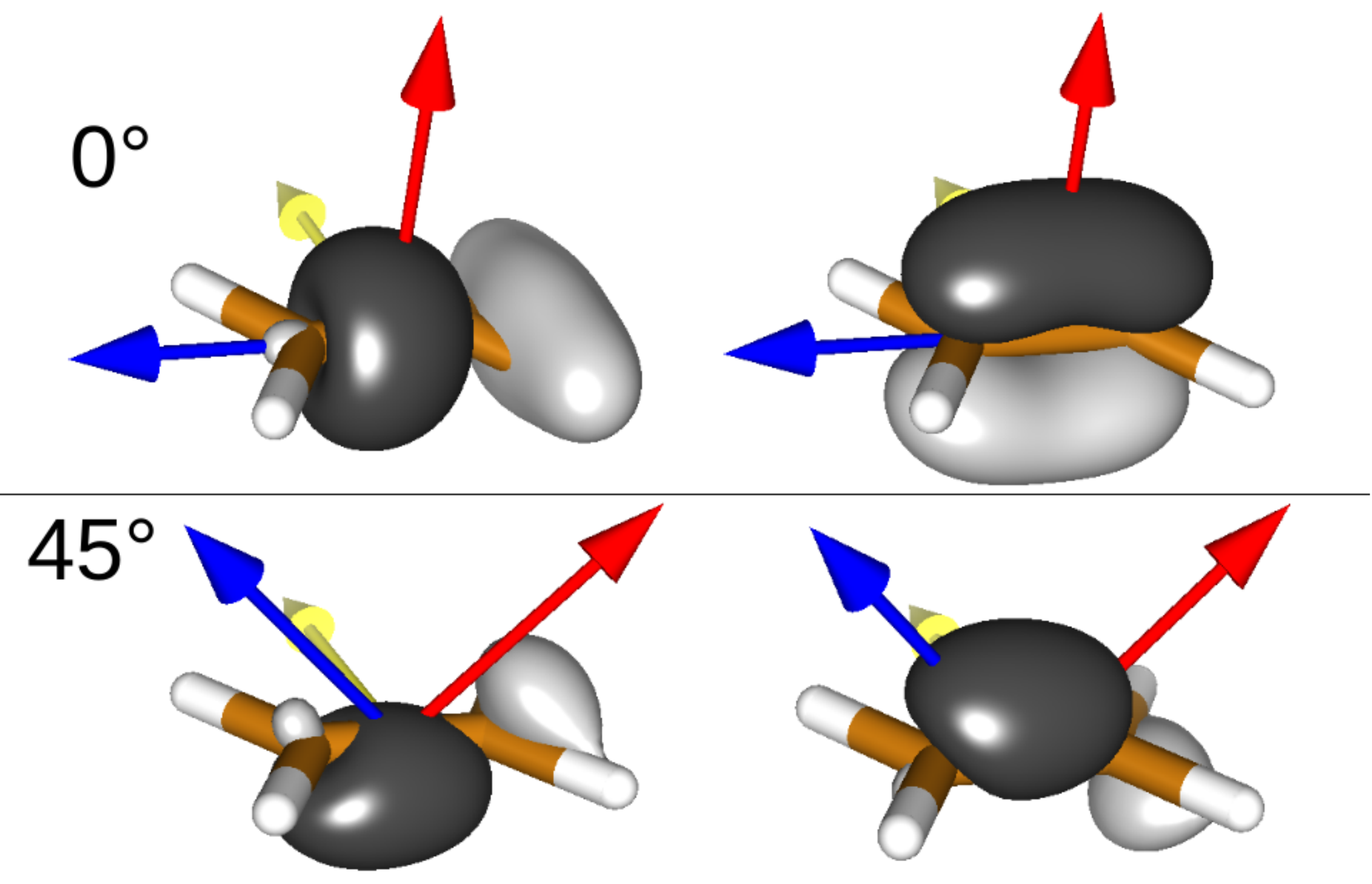}
    \caption{
    Graphical depiction of the two LMOs centered around the \ce{C=C} bond in ethylene (\ce{C2H4}) for two different molecular orientations generated using \SCDMM at the HF/cc-pVTZ level of theory.
    Since the basis functions are oriented with respect to the global coordinate system, failure to use the canonical/standard nuclear orientation may lead to PAOs that vary between these two molecular orientations.
    This directly results in \SCDMM LMOs with different character: when computed at $0^\circ$, the LMOs possess $\sigma$- and $\pi$-like character, while those computed at $45^\circ$ have more substantial $\tau$-like character.
    }
    \label{fig:angle2}
\end{figure}

To illustrate this theoretical dependence on the molecular orientation, we again consider the valence LMOs of ethylene (\ce{C2H2}) produced by the \SCDMM and \SCDML methods.
In particular, we will consider two molecular orientations of ethylene: (\textit{i}) $0^\circ$, in which the \ce{C=C} bond in ethylene is oriented along the $z$-axis (with the molecule in the $yz$ plane) and (\textit{ii}) $45^\circ$, in which the molecule has been rotated by $45^\circ$ towards the $x$-axis.
As expected, Fig.~\ref{fig:angle} shows that the locality of the \SCDMM and \SCDML LMOs does vary with respect to the molecular orientation, although there is significantly less variation observed in the \SCDML case.
Here, we again hypothesize that the symmetric orthogonalization procedure inherent to POAOs helps the \SCDML method ``recognize'' (and adapt to) the molecular orientation, thereby leading to more consistent orbital locality across the two orientations. 
Motivated by the changes observed in the locality of the \SCDMM LMOs in Fig.~\ref{fig:angle}, we also explored how the molecular orientation could impact the character of the \SCDMM LMOs centered around the \ce{C=C} bond in ethylene.
As depicted in Fig.~\ref{fig:angle2}, different molecular orientations can lead to substantial differences in the character of the LMOs, with the $0^\circ$ configuration yielding $\sigma$- and $\pi$-like orbitals and the $45^\circ$ configuration yielding $\tau$-like orbitals.
While this variation in orbital character is interesting, choosing a molecular orientation to induce a certain type of LMO is inadvisable; in situations where it is feasible, we recommend using the standard practice of reorienting the molecule into the canonical/standard nuclear orientation, which would avoid this dependence on the molecular orientation and ensure that the results are reproducible.

Here, we note in passing that localization methods like Boys and \SCDMG will also have a weak dependence on the molecular orientation through slight variations in the computed occupied space.
In the \SCDMG case, the use of atom-centered Treutler--Ahlrichs--Lebedev~\cite{tre95,leb1999} grids can potentially introduce an additional weak dependence on the molecular orientation since the origin employed for the angular (Lebedev) grid point discretization is again selected with respect to a global coordinate system.
As discussed in Sec.~\ref{sec:grid_appendix}, the output of \SCDMG is insensitive to small variations in these grid point locations provided that a sufficiently dense grid is used (\ie grid levels $\geq 4$).
While the use of the canonical/standard nuclear orientation also removes these dependencies for calculations involving a fixed molecule, these issues are important when dealing with evolving systems (\eg during MD simulations~\cite{iftimie_on-the-fly_2004}) where the molecule(s) must be allowed to translate and rotate with respect to the global coordinate system.
As such, the use of localization methods in this context is of both theoretical and practical interest, and remains the subject of ongoing research.

\end{document}